\documentclass[12pt,a4paper]{article}

\usepackage[T1]{fontenc}

\usepackage{mathrsfs}
\usepackage[francais,english]{babel}
\usepackage{mathpazo} 
\usepackage{amsfonts}
\usepackage{amssymb}
\usepackage{amsmath}
\usepackage{epsfig}
\usepackage{color}
\usepackage{graphicx}
\usepackage{nicefrac}
\usepackage[latin1]{inputenc}
\usepackage{slashed}
\usepackage{multirow}
\usepackage{comment}
\usepackage{soul}
\usepackage[nosort]{cite}
\usepackage{datetime}
\usepackage{braket}
\usepackage{bm}
\usepackage{cancel}
\usepackage{xcolor}

\usepackage[titletoc]{appendix}

\def\be{\begin{equation}}
\def\ee{\end{equation}}
\def\ba{\begin{eqnarray}}
\def\ea{\end{eqnarray}}

\def\del{\partial}

\def\c{\chi}

\def\s{\sigma}

\def\no{\noindent}

\def\qq{\qquad}

\def\IR{\relax{\rm I\kern-.18em R}}

\def\IR{\relax{\rm I\kern-.18em R}}
\def\IL{\relax{\rm I\kern-.18em L}}

\def\inv{^{\raise.15ex\hbox{${\scriptscriptstyle -}$}\kern-.05em 1}}

\def\Tr{{\rm Tr}}

\definecolor{lightred}{RGB}{255,127,127}
\definecolor{lightgreen}{RGB}{127,255,127}
\definecolor{lightblue}{RGB}{127,127,255}
\definecolor{linkcolor}{rgb}{0,0,0.6}
\usepackage[ pdftex,colorlinks=true,
pdfstartview=FitV,
linkcolor= linkcolor,
citecolor= linkcolor,
urlcolor= linkcolor,
hyperindex=true,
hyperfigures=false, linktocpage=true]
{hyperref}

\newcommand{\vp}{\varphi}
\newcommand{\p}{\partial}
\newcommand{\g}{\mathfrak{g}}
\newcommand{\dd}{\text{d}}
\newcommand{\W}[1]{I_{\text{W}\hspace{-1pt}\text{Z}}\bigl[#1\bigr]}
\newcommand{\Opm}[3]{\Omega^{#1\,#3}_{\;\;#2}}

\newcommand{\cg}{c_G}
\def\res{\mathop{\text{res}\,}}
\newcommand{\Ad}{\text{Ad}}
\newcommand{\ac}{a}

\DeclareFontEncoding{LS1}{}{}
\DeclareFontSubstitution{LS1}{stix}{m}{n}
\DeclareSymbolFont{stixsymbols}{LS1}{stixscr}{m}{n}
\SetSymbolFont{stixsymbols}{bold}{LS1}{stixscr}{b}{n}
\DeclareMathSymbol{\kay}{\mathalpha}{stixsymbols}{"6B}

\numberwithin{equation}{section}

\begin{document}

\begin{titlepage}
\begin{flushright}\small{ZMP-HH/20-20}
\end{flushright}
\begin{center}

\vskip .4 cm

\vskip .3 in

{\large\bf RG flows of integrable $\sigma$-models and the twist function}

\vskip 0.4in

{\bfseries Fran\c{c}ois Delduc}$^{a,}$\footnote{E-mail:~francois.delduc@ens-lyon.fr},\phantom{x}
{\bfseries Sylvain Lacroix}$^{b,}$\footnote{E-mail:~sylvain.lacroix@desy.de},\\
{\bfseries Konstantinos Sfetsos}$^{c,}$\footnote{E-mail:~ksfetsos@phys.uoa.gr}\phantom{x}
and\phantom{x} {\bfseries Konstantinos Siampos}$^{c,}$\footnote{E-mail:~konstantinos.siampos@phys.uoa.gr}

\vskip 0.35in

{\em${}^a$Univ Lyon, Ens de Lyon, Univ Claude Bernard,\\
CNRS, Laboratoire de Physique, F-69342 Lyon, France}

\vskip 0.1in

{\em${}^b$II. Institut f\"ur Theoretische Physik, Universit\"at Hamburg,\\
Luruper Chaussee 149, 22761 Hamburg, Germany\\
Zentrum f\"ur Mathematische Physik, Universit\"at Hamburg,\\
Bundesstrasse 55, 20146 Hamburg, Germany}

\vskip 0.1in

{\em${}^c$Department of Nuclear and Particle Physics,
Faculty of Physics,\\ National and Kapodistrian University
of Athens,\\15784 Athens, Greece}

\vskip 0.1in


\vskip .43in
\end{center}

\centerline{\bfseries Abstract}

\noindent
In the study of integrable non-linear $\sigma$-models which are assemblies and/or deformations of principal chiral models  and/or WZW models, 
a rational function called the twist function plays a central r\^ole. For a large class of such models, we show that they are one-loop renormalizable, and that the renormalization group flow equations can be written directly  in terms of the twist function in a remarkably simple way. The resulting equation appears to have a universal character when the integrable model is characterized by a twist function.

\noindent

\vskip .4in
\noindent
\end{titlepage}
\vfill
\eject

\newpage

\tableofcontents

\setlength{\parskip}{.5em}

 \setcounter{footnote}{0}
 
\section{Introduction}

In this article, we study the one-loop renormalization group flow of a class of integrable non-linear $\sigma$-models,  which in particular includes the integrable model coupling together an arbitrary number of principal chiral models with Wess--Zumino terms introduced in~\cite{Delduc:2018hty,Delduc:2019bcl}. Two methods are used, one based on known results relating the renormalization to geometrical data~\cite{Ecker:1972bm,Honerkamp:1971sh,Friedan:1980jf,Curtright:1984dz,Braaten:1985is,Fridling:1985hc}, the other one using heat-kernel techniques. 
The models under consideration are renormalizable in a field theoretical sense, namely the parameters entering the models flow under the renormalization group. When one takes into account this redefinition of the parameters, the form of the corresponding $\s$-model action stays the same under the renormalization group flow and no extra modifications are needed.

The models that we shall consider are integrable, in the sense that their field equations take the Lax form. This means that there exist matrices
${\mathscr L}_t(t,x;z)$ and ${\mathscr L}_x(t,x;z)$ taking value in some Lie algebra
$\g$ and depending on a complex spectral parameter $z$, which are such that the field equations can be deduced from the zero curvature equation
\begin{equation*}\partial_t{\mathscr L}_x-\partial_x{\mathscr L}_t-[{\mathscr L}_t,{\mathscr L}_x]=0\, .
\end{equation*}
This ensures that one can construct an infinite set of conserved charges. Moreover, in order for these charges to Poisson commute, the Poisson bracket of the Lax matrix ${\mathscr L}_x(t,x;z)$ has to take a particular form introduced by Maillet in~\cite{Maillet:1985,Maillet:1986}, which, in the tensor notation, reads
\begin{equation*}
\begin{split}
\big\{{\mathscr L}_{x\underline{1}}(t,x;z),{\mathscr L}_{x\underline{2}}(t,x';w)\big\}=
&\big(\big[r_{\underline{12}}(z,w),{\mathscr L}_{x\underline{1}}(t,x;z)]-[r_{\underline{21}}(w,z),{\mathscr L}_{x\underline{2}}(t,x';w)]\big)\delta(x-x')\cr
&\hspace{15pt} +\big(r_{\underline{12}}(z,w)+r_{\underline{21}}(w,z)\big)\partial_x\delta(x-x')\, .
\end{split}\end{equation*}
In all cases studied in this article, the $r$-matrix $r_{\underline{12}}(z,w)$,  taking value in $\g\otimes\g$ and satisfying the classical Yang--Baxter equation, takes a particular form
\begin{equation}r_{\underline{12}}(z,w)=\frac{C_{\underline{12}}}{z-w}\varphi^{-1}(w)\, ,
\label{Eq:rtwist}
\end{equation}
where $C_{\underline{12}}=\sum\limits_a T_a\otimes T_a$ is the tensor Casimir of the Lie algebra $\mathfrak{g}$ (with $T_a$'s an orthonormal basis of generators of $\mathfrak{g}$), while $\varphi(z)$ is a meromorphic function, known as the twist function. The pivotal r\^ole of the twist function was already stressed in other articles, see~\cite{Delduc:2013fga,Vicedo:2015}. As shown in~\cite{Vicedo:2017cge} the form \eqref{Eq:rtwist} of the $r$-matrix, and thus the existence of the twist function, is ensured provided the model under investigation may be interpreted as a realization of an affine Gaudin model, 
which is the case for the models that we consider in this article. 
In the context of $\lambda$-deformed theories, with prototype constructions in~\cite{Sfetsos:2013wia} and~\cite{Georgiou:2017jfi}, the r\^ole of the twist function has been elucidated in~\cite{{Hollowood:2015dpa}, Georgiou:2019plp}. Moreover, it has been shown recently~\cite{Costello:2019,Vicedo:2019dej,Delduc:2019c} that the twist function plays a crucial r\^ole in the four-dimensional Chern--Simons approach to integrable non-linear $\sigma$-models.

The twist function $\varphi(z)$ encodes all the continuous parameters of the models that we shall consider. Thus, instead of writing an renormalization group  flow equation for each of the parameters, it is tempting to try to write an equation for the twist function itself. Not only does this turn out to be possible, but moreover, it leads to a particularly compact and simple form of the renormalization group flow equation, namely
\begin{equation}\label{Eq:FlowTwist}
\frac{\dd\;}{\dd\tau} \vp(z) = - \cg \frac{\dd\;}{\dd z} \Big(\vp(z)f(z)\Big),
\end{equation}
where $\tau$ is related to the renormalization group  scale parameter  and $\cg$ is the value of the quadratic Casimir of $\g$ in the adjoint representation. The construction of the meromorphic function $f(z)$, tightly connected with the twist function, will be given in Section~\ref{Sec:Integrable}. This formula will be demonstrated in the case of the coupled principal chiral models with an arbitrary number of copies introduced in~\cite{Delduc:2018hty} and further studied in~\cite{Delduc:2019bcl}. 
However, we believe it to be widely applicable. Indeed, we check that it also holds for the cases of the $\lambda$-deformed model~\cite{Sfetsos:2013wia,Itsios:2014vfa,Itsios:2014x} and of the Yang--Baxter model with a Wess--Zumino term~\cite{Klimcik:2002zj,Klimcik:2008eq,Delduc:2014uaa,Squellari:2014,Sfetsos:2015nya,Demulder:2017zhz}, which are deformations of one copy of the principal chiral model. In addition,  we will show that this relation holds for the doubly $\lambda$-deformed models of~\cite{Georgiou:2017jfi}, which should be seen as a deformation of two copies of Wess--Zumino--Witten conformal field theories.

The article is organized as follows: In Section~\ref{Sec:CoupMod}, a general form of the action of the coupled models is given, which is not yet necessarily integrable. The geometry of the target manifold, its metric and $B$-field, is studied and the one loop renormalizability of this class of models is established using the Ricci tensor. In Section~\ref{Sec:Integrable}, the restriction to the integrable non-linear $\sigma$-models introduced in~\cite{Delduc:2018hty,Delduc:2019bcl} is given and their twist function is defined. Moreover, the geometry of these integrable models is studied. This requires the introduction of the function $f(z)$ appearing in \eqref{Eq:FlowTwist}. The renormalization of these coupled integrable models is studied in Section~\ref{Sec:RGInt}, where Equation~\eqref{Eq:FlowTwist} is obtained. The cases of the deformed models ($\lambda$-model, doubly $\lambda$-model and Yang--Baxter model) is studied in Section~\ref{Sec:deformed}. In Appendices~\ref{App:Ge} and~\ref{appendixheat} are gathered some facts about the renormalization of the coupled models, using geometric techniques and heat kernel techniques respectively.
In Appendix~\ref{Sec:2coupled} we study more closely the case where two copies of the principal chiral model are coupled: in particular, a family of models is studied, whose relation with the general framework is non trivial and is investigated in detail. The technical Appendices~\ref{App:GB},~\ref{App:SumR},~\ref{App:ROverRho} and~\ref{App:DiffDr} give the proofs of some important relations used in the main text.

\section{The general coupled \texorpdfstring{$\bm{\sigma}$}{sigma}-model on \texorpdfstring{$\bm{G^N}$}{GN}}
\label{Sec:CoupMod}

\subsection{Action, metric and antisymmetric tensor field}

Let $G$ be a compact Lie group and $g^{(1)}(x,t),\dots,g^{(N)}(x,t)$ be $G$-valued fields depending on two-dimensional space-time coordinates $(x,t)$. We will denote by $x^\pm = (t \pm x)/2$ the light-cone coordinates and by $\p_\pm=\p_t \pm \p_x$ the corresponding derivatives. We then define the light-cone Maurer--Cartan currents $j^{(i)}_\pm=g^{(i)-1}\p_\pm g^{(i)}$ or in terms of components
\begin{equation}
\label{leftMC}
j_\pm^{(i)a} = -\text{Tr}(T_ag^{(i)\,-1}\p_\pm g^{(i)})\, .
\end{equation}
Here, the $T_a$'s are anti-Hermitian matrices forming a basis of the Lie algebra $\g$ of $G$,
normalized as $\text{Tr}\left(T_aT_b\right)=-\delta_{ab}$ and obeying $[T_a,T_b]=f_{ab}{}^c\,T_c$,
where the structure constants $f_{ab}{}^c$ are real and the index $a$ runs from $1$ to dim $G$.
The $\g$-valued currents $j_\pm^{(i)}$ satisfy the Maurer--Cartan equation
\begin{equation}\label{Eq:Maurer}
\partial_+j_-^{(i)} - \partial_-j_+^{(i)} + \bigl[ j_+^{(i)}, j_-^{(i)} \bigr] =0\, .
\end{equation}
We consider the action
\begin{equation}
\label{Eq:Action}
S\bigl[g^{(1)},\dots,g^{(N)}\bigr] =- \sum_{i,j=1}^N \rho_{ij}\int \dd t \, \dd x\;
\text{Tr}(j^{(i)}_+ j^{(j)}_-) + \sum_{i=1}^N \kay_i \, \W {g^{(i)}},
\end{equation}
where the $\rho_{ij}$'s and $\kay_i$'s are constant parameters such that $4\pi \kay_i$ are integers.
The Wess--Zumino (WZ) term is given by
\begin{equation}
\label{action.WZ}
I_\text{WZ}[g^{(i)}]=\int\dd\xi\,\dd t\,\dd x\;\text{Tr}\left(g^{(i)-1}\partial_\xi g^{(i)}[g^{(i)-1}\partial_t g^{(i)},g^{(i)-1}\partial_x g^{(i)}]\right)\,.
\end{equation}
The action \eqref{Eq:Action} is invariant under the parity transformation
\begin{equation}
\label{parity}
x^\pm\to x^\mp\,,\qquad \rho_{ij}\to \rho_{ji}\,,\qquad \kay_i\to - \kay_i\,.
\end{equation}
The field equation coming from a variation of $g^{(i)}$ reads
\begin{equation}\label{Eq:FieldEq}
\sum_{j=1}^N\Big(\rho_{ij}\bigl(\partial_+j_-^{(j)}+\bigl[j_+^{(i)},j_-^{(j)}\bigr] \bigr)+\rho_{ji}\bigl(\partial_-j_+^{(j)}+\bigl[j_-^{(i)},j_+^{(j)}\bigr]\bigr)\Big)
+\frac{\kay_i}{2}\bigl(\partial_+j_-^{(i)}-\partial_-j_+^{(i)}\bigr)=0\, .
\end{equation}
We define the ``metric'' and the ``$B$-field'' of the model as the following symmetric and 
anti-symmetric tensors\footnote{More precisely, the quantities $G_{ij}$ and $B_{ij}$ are the parts of the metric and the $B$-field which describe the interaction between the $N$ different copies of the group $G$. The complete metric and $B$-field also contain parts describing the internal geometry of each copy of $G$ as well as the WZ terms (see Appendix~\ref{App:Ge} for more details and in particular \eqref{Eq:FullMetB}). By a slight abuse of language, we will also call $G_{ij}$ and $B_{ij}$ metric and $B$-field.}
\begin{equation}
\label{Eq:MetB}
G_{ij} = \frac{\rho_{ij}+\rho_{ji}}{2} \;\;\;\;\; \text{ and } \;\;\;\;\; B_{ij} = \frac{\rho_{ij}-\rho_{ji}}{2}\, .
\end{equation}
We will denote by $G^{ij}$ the inverse of the metric $G_{ij}$ and will lower and raise indices with the metric $G_{ij}$ and the inverse metric $G^{ij}$, respectively.
For future convenience, we also define the coefficients
\begin{equation}\label{Eq:R}
r^+_i = \sum_{j=1}^N \rho_{ij} - \frac{\kay_i}{2} \;\;\;\; \text{ and } \;\;\;\; r^-_i = \sum_{j=1}^N \rho_{ji} + \frac{\kay_i}{2}\, .
\end{equation}

\subsection{Renormalization and Ricci tensor}
\label{SubSec:RGGeneral}

Using the standard geometric techniques developed in~\cite{Ecker:1972bm,Honerkamp:1971sh,Friedan:1980jf,Curtright:1984dz,Braaten:1985is,Fridling:1985hc}, we prove in Appendix~\ref{App:Ge} that the $\sigma$-model \eqref{Eq:Action} is one-loop renormalizable.
Alternatively, this result can be obtained using heat kernel techniques, as we prove in Appendix~\ref{appendixheat}.
More precisely, we find that the corresponding renormalization group flow of the coefficients $\rho_{ij}$ and $\kay_i$ is given by
\begin{equation}
\label{Eq:RG}
\frac{\dd\rho_{ij}}{\dd\tau} = R^+_{ij} \;\;\;\;\; \text{ and } \;\;\;\;\; \frac{\dd\kay_i}{\dd\tau} = 0\,,
\end{equation}
where $\tau=\frac{1}{8\pi}\ln\mu^2$ and $\mu$ is the energy scale.
In this expression, $R^+_{ij}$ denotes the torsionfull Ricci tensor of the model.\footnote{More precisely, $R^+_{ij}$ is the main part of the torsionfull Ricci tensor of the model (see Appendix~\ref{App:Ge} for more details and in particular \eqref{Eq:RFull}). We will still call it Ricci tensor by a slight abuse of language.}
This tensor reads
\begin{equation}
\label{RGfinal}
 R^+_{ij} =
\cg \sum_{k,l=1}^N \omega^{-k}{}_{\;\, li} \omega^{+l}{}_{\;\, kj}
\, ,
\end{equation}
where $\cg$ is the value of the quadratic Casimir in the adjoint representation, \textit{i.e.}
${f_{db}}^c{f_{ca}}^d = - \cg\, \delta_{ab}$, and $\omega^{\pm i}{}_{jk}$ are the torsionfull spin connections
\be
\label{Eq:Omega}
\begin{split}
\omega^{+l}{}_{\;\, kj} &= \frac{1}{2}(G^{lk}-G^{lj}) \rho_{kj} - \frac{\kay_j}{4}G^{lk}\delta_{kj} + \frac{\delta_{kj}}{2}  \rho_j{}^l\, , 
\\
\omega^{-k}{}_{\;\, li} &= \frac{1}{2}(G^{kl}- G^{ki})\rho_{il} + \frac{\kay_i}{4}G^{kl}\delta_{li} + \frac{\delta_{li}}{2}  \rho^k{}_i\, .
\end{split}
\ee
These two connections are related by the generalized parity transformation given in \eqref{parity}. We note that the fact that the $\kay_i$'s are not flowing in Equation \eqref{Eq:RG} is consistent with the topological nature of the WZ terms.

\section{The integrable models}
\label{Sec:Integrable}

\subsection{Definitions and twist function}

\paragraph{Defining parameters and action:} The integrable model considered in~\cite{Delduc:2018hty,Delduc:2019bcl} corresponds
to a particular choice for the coefficients $\rho_{ij}$ and $\kay_i$ in the action \eqref{Eq:Action}, expressed in terms of $3N$ defining parameters $z_i$ and $\zeta_i^\pm$.
Let us define the functions
\begin{equation}\label{Eq:PhiPm}
\vp_\pm(z) = \prod_{i=1}^N \frac{z-\zeta_i^\pm}{z-z_i} \;\;\;\; \text{ and } \;\;\;\; \vp_{\pm,i}(z) = (z-z_i)\vp_\pm(z)\, ,
\end{equation}
where in particular we note  that $\vp_{\pm,i}(z)$ is regular at $z=z_i$. The model of~\cite{Delduc:2018hty,Delduc:2019bcl} then corresponds to the following choice for the coefficients $\rho_{ij}$ and $\kay_i$:\footnote{\label{FootnoteDilation}Note that in the notation of~\cite{Delduc:2018hty,Delduc:2019bcl}, we choose $\ell^\infty=1$, which one can always do without loss of generality using the freedom of rescaling the spectral parameter $z\mapsto \gamma z$.}
\begin{subequations}
\label{Eq:RhoK}
\begin{eqnarray}
\rho_{ii} &= & \frac{1}{4} \Bigl( \vp_{+,i}'(z_i)\vp_{-,i}(z_i) - \vp_{+,i}(z_i)\vp_{-,i}'(z_i) \Bigr)\, , 
\\
\rho_{ij} &= & \frac{1}{2} \frac{\vp_{+,i}(z_i)\vp_{-,j}(z_j)}{z_i-z_j}\, , \;\;\; \text{ if } \;\; i\neq j\, ,
\label{Eq:Rhoij} \\
\kay_i &= & \frac{1}{2} \Bigl( \vp_{+,i}'(z_i)\vp_{-,i}(z_i) + \vp_{+,i}(z_i)\vp_{-,i}'(z_i) \Bigr)\, . 
\label{Eq:K}
\end{eqnarray}
\end{subequations}
The expressions in \eqref{Eq:RhoK} are invariant under the transformation 
\begin{equation}
\label{glool}
z_i \mapsto z_i+a\,,\qquad \zeta_i^\pm \mapsto \zeta_i^\pm + a\, ,
\end{equation}
of the $3N$ defining parameters $(z_i, \zeta_i^\pm)$ of the model.
Thus, the model effectively depends only on $3N-1$ free parameters.
This redundancy in shifting the parameters $z_i$ and $\zeta_i^\pm$ can be fixed by setting one of them to a specific value. However, to keep the discussion as symmetric as possible, we will keep all these parameters free in the rest of this article.
Finally, let us note that the parity transformation \eqref{parity} acts on $(z_i,\zeta^\pm_i)$ as
\begin{equation}
z_i\to-z_i\,,\qquad \zeta^\pm_i  \to - \zeta^\mp_i\, .
\end{equation}
Using the identity
\begin{equation}
\vp'_{\pm,i}(z_i) = 1 + \sum_{j\neq i=1}^N \frac{\vp_{\pm,j}(z_j)}{z_i-z_j}\, ,
\end{equation}
one sees that the coefficients $r^\pm_i$  in \eqref{Eq:R} take a very simple form in the integrable case
\begin{equation}\label{Eq:RPhi}
r^\pm_i = \mp \frac{1}{2} \vp_{\pm,i}(z_i)\, .
\end{equation}
One can then rewrite \eqref{Eq:Rhoij} as
\begin{equation}\label{Eq:RhoR}
\rho_{ij} = -2\frac{r_i^+\, r_j^-}{z_i - z_j}\, , \;\;\;\;\;\; \text{ for } \; i \neq j\, ,
\end{equation}
or equivalently as
\begin{equation}\label{Eq:DiffZ}
z_i - z_j = -2\frac{r_i^+\, r_j^-}{\rho_{ij}}\, , \;\;\;\;\;\; \text{ for } \; i \neq j\, .
\end{equation}

\paragraph{Twist function:}The model defined by the coefficients $\rho_{ij}$ and $\kay_i$ given in  \eqref{Eq:RhoK} is classically integrable. In particular, it possesses a Lax matrix whose Poisson brackets assume the form of a Maillet bracket with twist function. The latter is simply defined as
\begin{equation}\label{Eq:Twist}
\vp(z) = - \vp_+(z) \vp_-(z) = -\prod_{i=1}^N \frac{(z-\zeta_i^+)(z-\zeta_i^-)}{(z-z_i)^2}\, .
\end{equation}
The partial fraction decomposition of the twist function is given by
\begin{equation}
\label{Eq:TwistLevels}
\vp(z) = \sum_{i=1}^N \left( \frac{\ell_i}{(z-z_i)^2} - \frac{2\kay_i}{z-z_i} \right) - 1\, ,
\end{equation}
where the coefficient $\kay_i$ is defined by \eqref{Eq:K} and $\ell_i$ by
\begin{equation}\label{Eq:LR}
\ell_i =- \vp_{+,i}(z_i) \vp_{-,i}(z_i) = \frac{1}{4} r_i^+ r_i^-\, .
\end{equation}
Let us note that the inverse of the twist function can be written as
\begin{equation}\label{Eq:TwistInv}
\frac{1}{\vp(z)} = \sum_{i=1}^{N} \frac{1}{\vp'(\zeta^+_i)} \frac{1}{z-\zeta^+_i} + \sum_{i=1}^{N} \frac{1}{\vp'(\zeta^-_i)} \frac{1}{z-\zeta^-_i} - 1\, .
\end{equation}
Furthermore, we introduce the functions
\begin{equation}
\label{Eq:fpm}
f_\pm(z) = \sum_{i=1}^{N} \frac{1}{\vp'(\zeta^\pm_i)} \frac{1}{z-\zeta^\pm_i} - \frac{1}{2} \pm \frac{\ac}{2}\, ,
\end{equation}
where $\ac$ is an arbitrary constant.
In terms of these, one can rewrite the inverse \eqref{Eq:TwistInv} of the twist function and introduce a new function $f(z)$ through
\begin{equation}
\label{Eq:SumDiffFpm}
\frac{1}{\vp(z)} = f_+(z) + f_-(z) \qquad \text{ and } \qquad f(z) =  f_+(z) - f_-(z)\, .
\end{equation}
The inverse of the twist function possesses a double zero at $z_i$. Moreover, one shows from \eqref{Eq:TwistLevels} that its second derivative at $z_i$ is equal to $\nicefrac{2}{\ell_i}$. We then have
\begin{equation}
f_+(z_i) + f_-(z_i) = 0\, , \;\;\;\;\;\;\; f'_+(z_i) + f'_-(z_i) = 0 \;\;\;\;\;\;\; \text{and} \;\;\;\;\;\;\; f''_+(z_i) + f''_-(z_i) = \frac{2}{\ell_i}\, .
\end{equation}
Thus, we also have
\begin{equation}
\label{Eq:FpmId}
f_\pm(z_i) = \pm \frac{1}{2} f(z_i)\, , \;\;\;\;\;\;\;\; f'_\pm(z_i) = \pm \frac{1}{2} f'(z_i) \;\;\;\;\;\;\;\; \text{and} \;\;\;\;\;\;\;\;\; f''_\pm(z_i) = \pm \frac{1}{2} f''(z_i) + \frac{1}{\ell_i}\, .
\end{equation}

\subsection{Geometry of the model}
\label{SubSec:Geom}

When the coefficients $\rho_{ij}$ and $\kay_i$ are given by their integrable expressions \eqref{Eq:RhoK}, the geometry of the underlying target space $G^N$ satisfies various useful properties that we summarize below. For brevity, we gather the proofs of these results in the Appendices~\ref{App:GB},~\ref{App:SumR} and~\ref{App:ROverRho}.

\paragraph{Inverse metric and $\bm{B}$-field:} The inverse metric $G^{ij}$ of the integrable coupled model can be expressed as (see Appendix~\ref{App:GB})
\begin{equation}\label{Eq:GInv}
G^{ii} = 2f'(z_i) \qquad \text{ and } \qquad G^{ij} = 2 \frac{f(z_i)-f(z_j)}{z_i-z_j}\ , \;\; \text{ if } i\neq j\,.
\end{equation}
Clearly, this expression is independent of the parameter $\ac$ introduced in \eqref{Eq:SumDiffFpm}. Note that the expression for $G^{ii}$ can be obtained by taking the limit $z_j\to z_i$ in the expression for $G^{ij}$. 

\no
Using Equation~\eqref{Eq:GInv}, one can derive some useful identities obeyed by the inverse metric $G^{ij}$. In particular, one has
\begin{equation}\label{Eq:GInvSym}
\frac{G^{ij}-G^{ik}}{z_j-z_k} = \frac{G^{ji}-G^{jk}}{z_i-z_k} \, ,
\end{equation}
for every $i,j,k \in \lbrace 1,\dots,N \rbrace$ with $k$ distinct from $i$ and $j$.

\no
When the coefficients $\rho_{ij}$ and $\kay_i$ take their integrable form \eqref{Eq:RhoK}, the $B$-field 
defined in \eqref{Eq:MetB} satisfies (see Appendix~\ref{App:GB}) 
\begin{equation}
\begin{split}
\label{Eq:B}
&B^i_{\;i} =  \dfrac{\ell_i}{2} f''(z_i) - \kay_i f'(z_i)\, ,
 \\
&B^i_{\;j} =  \dfrac{f(z_i)-f(z_j)}{z_i-z_j} \left(\frac{\ell_j}{z_i-z_j} - \kay_j \right) - \frac{\ell_j f'(z_j)}{z_i-z_j}\, ,
 \;\;\; \text{ if } i\neq j\, , 
\end{split}
\end{equation}
where $\ell_i$ is defined as in \eqref{Eq:LR} and indices are raised using $G^{ij}$. 
Let us note that combining equations \eqref{Eq:GInv} and \eqref{Eq:B}, one obtains the following identity, for $i$ and $j$ distinct:
\begin{equation}\label{Eq:GInvToH}
\frac{\ell_j}{2}\frac{G^{ij}-G^{jj}}{z_i-z_j} = B^i_{\;j} + \frac{\kay_j}{2} G^{ij}\, .
\end{equation}

\paragraph{Ricci tensor:} 
The Ricci tensor $R^+_{ij}$ of the integrable model, given by  \eqref{RGfinal}, also satisfies a certain number of identities. For instance, one has (see Appendix~\ref{App:SumR})
\begin{equation}\label{Eq:SumR}
\sum_{j=1}^N R^+_{ij} = \Lambda_i^+ r^+_i \;\;\;\;\;\; \text{ and } \;\;\;\;\;\; \sum_{j=1}^N R^+_{ji} = \Lambda_i^- r^-_i\, ,
\end{equation}
where
\begin{equation}\label{Eq:Lambda}
\Lambda^\pm_i = \frac{\cg}{4} \, G^{ii} \pm \frac{\cg}{8} \sum_{j=1}^N \left( 4G^{ij} B^j_{\;j} + 2 G^{jj}B^i_{\;j} + \kay_j G^{ij}G^{jj} \right).
\end{equation}
Moreover, for $i\neq j$, one has (see Appendix~\ref{App:ROverRho})
\begin{equation}\label{Eq:ROverRho}
R^+_{ij} =\rho_{ij}\Big( \Lambda^+_i + \Lambda^-_j- \frac{\cg}{2} G^{ij}\Big)\, .
\end{equation}

\section{Renormalization of the integrable coupled model}
\label{Sec:RGInt}

We stated in Subsection~\ref{SubSec:RGGeneral} that the general coupled $\sigma$-model on $G^N$ defined by the action \eqref{Eq:Action} is renormalizable at one-loop -- the proof is in Appendix~\ref{App:Ge}. Moreover, we determined the corresponding renormalization group (RG) flow \eqref{Eq:RG} of the coefficients $\rho_{ij}$ and $\kay_i$ appearing in this action.

It is not obvious that the integrable model described in Section~\ref{Sec:Integrable} is one-loop renormalizable.
Indeed, it corresponds to a particular restriction of \eqref{Eq:Action} and the RG flow might not respect this restriction.
More precisely, recall that this model corresponds to a particular choice \eqref{Eq:RhoK} of the coefficients $\rho_{ij}$ and $\kay_i$, in terms of $3N$ defining parameters $z_i$ and $\zeta_i^\pm$, $i\in\lbrace 1,\dots,N\rbrace$. 
Hence, it is important to show that this choice is stable under the RG flow, and thus that renormalization just leads to a flow of the defining parameters $(z_i,\zeta_i^\pm)$.
Indeed, using the general expression \eqref{Eq:RG} of the RG flow of the coefficients $\rho_{ij}$ and $\kay_i$, combined with various identities obeyed by the metric, $B$-field and Ricci tensor of the integrable model (see previous section), we will extract in Subsection~\ref{SubSec:RGParam} the induced RG flow on the defining parameters $z_i$ and $\zeta_i^\pm$ of this model (or equivalently, in a more compact way, on its twist function $\vp(z)$). To conclude the proof of the renormalizability of the model, we will then prove in Subsection~\ref{SubSec:RGConsistent} that, conversely, the corresponding flow of $z_i$ and $\zeta_i^\pm$ unambiguously leads to the flow \eqref{Eq:RG} of the coefficients $\rho_{ij}$ and $\kay_i$.

As a complement of this section, we study in more details the models coupling together two copies of the principal chiral model (PCM)
with WZ term in Appendix \ref{Sec:2coupled}. In particular, we investigate the renormalization and the integrability of a certain family of such models which does not directly fit in the class considered in Section \ref{Sec:Integrable}.

\subsection{RG flow of the defining parameters of the model}
\label{SubSec:RGParam}

\paragraph{Flow of the coefficients $\bm{r^\pm_i}$:} In this subsection, we compute the RG flow of the defining parameters $z_i$ and $\zeta_i^\pm$ of the integrable coupled $\sigma$-model. For that, we will need various intermediate steps, starting with the RG flow of the coefficients $r^\pm_i$ defined in \eqref{Eq:R}. Using \eqref{Eq:RG} for the RG flow of the coefficients $\rho_{ij}$ and $\kay_i$, we get
\begin{equation}
\frac{\dd r_i^+ }{\dd\tau} = \sum_{j=1}^N R^+_{ij} \qquad \text{ and } \qquad \frac{\dd r_i^-}{\dd\tau} = \sum_{j=1}^N R^+_{ji}\, .
\end{equation}
Using the identity \eqref{Eq:SumR} obeyed by the Ricci tensor of the integrable model, we then simply write the RG-flow of the coefficients $r^\pm_i$ as
\begin{equation}\label{Eq:FlowRpm}
\frac{\dd r_i^\pm }{\dd\tau} = \Lambda^\pm_i \, r^\pm_i.
\end{equation}

\paragraph{Flow of the  poles $\bm{z_i}$:} Let us consider $i,j\in\lbrace 1,\dots,N\rbrace$ distinct. Taking the logarithmic derivative of \eqref{Eq:DiffZ} with respect to $\tau$, one gets
\begin{equation}
\frac{1}{z_i-z_j} \frac{\dd(z_i-z_j)}{\dd\tau} = \frac{1}{r_i^+} \frac{\dd  r_i^+}{\dd\tau} + \frac{1}{r_i^-} \frac{\dd  r_i^-}{\dd\tau} - \frac{1}{\rho_{ij}} \frac{\dd\rho_{ij}}{\dd\tau}\, .
\end{equation}
Using \eqref{Eq:RG} and \eqref{Eq:FlowRpm} for the RG flow of the coefficients $\rho_{ij}$ and $r^\pm_i$, as well as  \eqref{Eq:ROverRho}, we get
\begin{equation}
\label{Eq:FlowDiffZ}
\frac{1}{z_i-z_j} \frac{\dd(z_i-z_j)}{\dd\tau} = \frac{\cg}{2}G^{ij} = \cg \frac{f(z_i)-f(z_j)}{z_i-z_j}\, ,
\end{equation}
where the last equality follows from \eqref{Eq:GInv}. One then extracts the RG flow of the parameters $z_i$ as
\begin{equation} 
\label{Eq:FlowZ}
\boxed{\frac{\dd z_i}{\dd\tau} = \cg\,f(z_i) \, .}
\end{equation}
Actually, there could be an additional additive term on the right hand side of \eqref{Eq:FlowZ}, the same for all values of the index $i$, which generically would be an arbitrary function of the $z_i$'s and $\zeta^\pm_i$'s. This reflects the fact that the model is invariant under a global translation \eqref{glool} of the parameters.
Such a term in the flow of $z_i$'s can always be reabsorbed in the coefficient $\ac$ introduced in the definition \eqref{Eq:fpm} of the functions $f_\pm(z)$.

\paragraph{Flow of the levels $\bm{\ell_i}$:} Recall the definition \eqref{Eq:LR} of the levels $\ell_i$ of the model. Using the expression \eqref{Eq:FlowRpm} of the RG flow of the coefficients $r^\pm_i$, one finds that
\begin{equation}
\label{Eq:FlowL}
\frac{\dd \ell_i}{\dd\tau} = (\Lambda^+_i+\Lambda^-_i)\ell_i = \frac{\cg }{2}\,G^{ii}\, \ell_i 
= \cg\,f'(z_i)\, \ell_i\, ,
\end{equation}
where in the second equality we used \eqref{Eq:Lambda} for the coefficients $\Lambda^\pm_i$ and in the last one
we used \eqref{Eq:GInv}.

\paragraph{Flow of the twist function:} Recall the expression \eqref{Eq:TwistLevels} of the twist function $\vp(z)$ of the model. Its RG flow is given by
\begin{equation}\label{Eq:FlowTwistLevels}
\frac{\dd\;}{\dd\tau}\vp(z) = \sum_{i=1}^N \left[ \frac{2\ell_i}{(z-z_i)^3} \frac{\dd z_i}{\dd \tau} +
\frac{1}{(z-z_i)^2} \left( \frac{\dd\ell_i}{\dd \tau} -2\kay_i \frac{\dd z_i}{\dd \tau} \right)\right]\, ,
\end{equation}
where we used the fact that the RG flow \eqref{Eq:RG} leaves the WZ levels $\kay_i$ invariant. 
Using \eqref{Eq:FlowZ} and \eqref{Eq:FlowL} yields
\begin{equation}\label{Eq:FlowTwist2}
\frac{\dd\;}{\dd\tau}\vp(z) = \cg \sum_{i=1}^N \left( \frac{2f(z_i)\ell_i}{(z-z_i)^3} + \frac{f'(z_i)\ell_i-2f(z_i)\kay_i}{(z-z_i)^2} \right)\, .
\end{equation}
Let us show that this can be re-expressed in a compact way. We consider the function $\vp(z)f(z)$. From the definition \eqref{Eq:SumDiffFpm} of $f(z)$, one sees that $f(z)$ has simple poles at the zeroes $\zeta^\pm_i$ of $\vp(z)$. Thus the function $\vp(z)f(z)$ is regular at the $\zeta^\pm_i$'s. Its poles are then only situated at the poles $z_i$ of $\vp(z)$ (and have multiplicity two). Note finally that $\vp(z)f(z)$ tends to $-\ac$ when $z$ goes to infinity.
The partial fraction decomposition of $\vp(z)f(z)$ is then of the form
\begin{equation}
\vp(z) f(z) = \sum_{i=1}^N \left( \frac{\gamma_i}{(z-z_i)^2} + \frac{\delta_i}{z-z_i} \right) - \ac\, ,
\end{equation}
for some coefficients $\gamma_i$ and $\delta_i$. The coefficient $\gamma_i$ is found to be
\begin{equation}
\gamma_i = \lim_{z\to z_i} \left((z-z_i)^2 \vp(z) f(z)\right) = f(z_i) \ell_i\, .
\end{equation}
Similarly, one can compute the coefficient $\delta_i$ using
\begin{equation*}
\vp(z) = \frac{\ell_i}{(z-z_i)^2} - \frac{2\kay_i}{z-z_i} + O\bigl((z-z_i)^0\bigr) \, ,
\quad f(z) = f(z_i) + f'(z_i) (z-z_i) + O\bigl( (z-z_i)^2 \bigr)\, ,
\end{equation*}
yielding
\begin{equation}
\delta_i = f'(z_i) \ell_i - 2 f(z_i)\kay_i\, .
\end{equation}
Thus, we have
\begin{equation}
\vp(z) f(z) = \sum_{i=1}^N \left( \frac{f(z_i)\ell_i}{(z-z_i)^2} + \frac{f'(z_i)\ell_i-2f(z_i)\kay_i}{z-z_i} \right) - \ac\, .
\end{equation}
The RG flow \eqref{Eq:FlowTwist2} of the twist function can then be rewritten as  in equation \eqref{Eq:FlowTwist} given in the introduction, restated here for the reader's convenience:
\begin{equation}\label{Eq:FlowTwist3}
\boxed{\frac{\dd\;}{\dd\tau} \vp(z) = - \cg \frac{\dd\;}{\dd z} \Big(\vp(z)f(z)\Big).}
\end{equation}
This encodes in a remarkably compact and simple way the RG flow of all the parameters of the model.

\paragraph{Flow of the zeroes $\bm{\zeta_i^\pm}$:} The RG flow  of the poles $z_i$ was determined in \eqref{Eq:FlowZ}. 
The remaining set of defining parameters of the integrable models are the zeroes $\zeta_i^\pm$ of the 
twist function, to which we now turn our attention.
From the factorized expression \eqref{Eq:Twist} of $\vp(z)$, one gets
\begin{equation}
\frac{1}{\vp(z)} \frac{\dd\;}{\dd\tau}\vp(z) = \sum_{i=1}^N \left(  \frac{2}{z-z_i} \frac{\dd z_i}{\dd\tau} - \frac{1}{z-\zeta^+_i} \frac{\dd \zeta_i^+}{\dd\tau} - \frac{1}{z-\zeta^-_i} \frac{\dd \zeta_i^-}{\dd\tau} \right).
\end{equation}
We then have that
\begin{equation}\label{Eq:ExtractFlowZeta}
\frac{\dd\zeta_i^\pm}{\dd\tau} = - \res_{z=\zeta_i^\pm}  \left( \frac{1}{\vp(z)} \frac{\dd\;}{\dd\tau}\vp(z) \right) =
\cg \res_{z=\zeta_i^\pm}  \left( \frac{1}{\vp(z)} \frac{\dd\;}{\dd z}\left(\vp(z)f(z)\right) \right),
\end{equation}
where in the last equality we have used \eqref{Eq:FlowTwist3}. 
The computation of this residue is made difficult by the fact that $f(z)$ has a pole at $\zeta_i^\pm$. To overcome this difficulty, let us recall the functions $f_+(z)$ and $f_-(z)$ introduced in \eqref{Eq:fpm} and note in particular that $f_\mp(z)$ is regular at $z=\zeta_i^\pm$. We will now explain how to use this fact to compute the above residue, focusing for simplicity on the case of $\zeta_i^+$. From \eqref{Eq:SumDiffFpm} we get that
\begin{equation}
\vp(z)f(z) = \vp(z) \bigl( f_+(z) - f_-(z) \bigr) = \vp(z) \bigl( f_+(z) + f_-(z) - 2f_-(z) \bigr) = 1 - 2 \vp(z)f_-(z).
\end{equation}
Thus, we have
\begin{equation}
\res_{z=\zeta_i^+}  \left( \frac{1}{\vp(z)} \frac{\dd\;}{\dd z}\left(\vp(z)f(z)\right) \right) = -2 \res_{z=\zeta_i^+}  \left( \frac{\vp'(z)}{\vp(z)} f_-(z) + f'_-(z) \right).
\end{equation}
Using the fact that $f_-(z)$ and $f'_-(z)$ are regular at $z=\zeta_i^+$ and the fact that the logarithmic derivative $\vp'(z)/\vp(z)$ has a simple pole at $z=\zeta_i^+$ with residue $1$, the above expression evaluates to $-2 \,f_-(\zeta^+_i)$. A similar computation yields the residue at $\zeta_i^-$ in terms of the function $f_+(z)$. In the end, one obtains
\begin{equation}\label{Eq:ResZeta}
\res_{z=\zeta_i^\pm}  \left( \frac{1}{\vp(z)} \frac{\dd\;}{\dd z}\left(\vp(z)f(z)\right) \right) = \mp 2 \,f_\mp(\zeta^\pm_i)\, .
\end{equation}
Re-inserting this result in \eqref{Eq:ExtractFlowZeta}, we obtain the RG flow of the zeroes $\zeta_i^\pm$
\begin{equation}
\label{rgzeta}
\boxed{\frac{\dd\zeta_i^\pm}{\dd\tau} = \mp 2\cg \, f_\mp(\zeta^\pm_i)\, .}
\end{equation}

\subsection{Consistency of the RG flow}
\label{SubSec:RGConsistent}

In the previous subsection, we showed that the RG flow \eqref{Eq:RG} of the coefficients $\rho_{ij}$ and $\kay_i$
induces a flow on the defining parameters $z_i$ and $\zeta_i^\pm$ of the model given by \eqref{Eq:FlowZ} and \eqref{rgzeta}, which we summarise here for the reader's convenience:
\begin{equation}\label{Eq:FlowDefParam}
\frac{\dd z_i}{\dd\tau} = \cg\,f(z_i) = \mp 2\cg\,f_\mp(z_i) \;\;\;\;\;\;\; \text{ and } \;\;\;\;\;\;\; \frac{\dd\zeta_i^\pm}{\dd\tau} = \mp 2\cg \, f_\mp(\zeta^\pm_i)\, ,
\end{equation}
where the second equality in the first equation comes from the identity \eqref{Eq:FpmId}.
To achieve the proof of the renormalizability of the model, we need to show that the flow is consistent, \textit{i.e.} that reinserting \eqref{Eq:FlowDefParam} in the expression \eqref{Eq:RhoK} of the coefficients $\rho_{ij}$ and $\kay_i$ in terms of the $z_i$'s and $\zeta_i^\pm$'s gives back \eqref{Eq:RG}.
For that, we will mostly follow the steps of the previous subsection in the reverse order. Although we will use some of the identities proved in the previous subsection, we will not use the various RG flow equations derived there, since our goal here is to rederive them starting from \eqref{Eq:FlowDefParam}.

\paragraph{Flow of the twist function $\bm{\vp(z)}$ and levels $\bm{\kay_i}$ and $\bm{\ell_i}$:} Let us thus start by proving that the flow \eqref{Eq:FlowDefParam} induces the flow \eqref{Eq:FlowTwist2} on $\vp(z)$. Using the expression \eqref{Eq:Twist} of $\vp(z)$ as a product, the flow of its logarithm can be written as
{\small \begin{equation}\label{Eq:LogFlowTwist}
\begin{split}
\frac{1}{\vp(z)} \frac{\dd\;}{\dd\tau} \vp(z) & = \sum_{i=1}^N \left( 2\frac{\dd z_i}{\dd\tau} \frac{1}{z-z_i} - \frac{\dd \zeta_i^+}{\dd\tau} \frac{1}{z-\zeta_i^+} - \frac{\dd \zeta_i^-}{\dd\tau} \frac{1}{z-\zeta_i^-} \right) \\
&= 2\cg \sum_{i=1}^N \left( \frac{f(z_i)}{z-z_i} + \frac{f_-(\zeta^+_i)}{z-\zeta_i^+} - \frac{f_+(\zeta_i^-)}{z-\zeta_i^-} \right).
\end{split}
\end{equation}}
To prove that this induces \eqref{Eq:FlowTwist2}, we thus have to show that the  last line coincides with
\begin{equation}
-\frac{\cg}{\vp(z)} \frac{\dd\;}{\dd z}\left(\vp(z)f(z)\right) = -\cg \frac{\vp'(z)}{\vp(z)} f(z) - \cg f'(z)\, .
\end{equation}
This function has poles at the points $z_i$ and $\zeta_i^\pm$, which one can show are simple. 
Using the fact that, at $z=z_i$, $f(z)$ and $f'(z)$ are regular and $\vp'(z)/\vp(z)$ has residue $-2$, one sees that the residue of this function at $z=z_i$ is $2\cg f(z_i)$. Moreover, we showed in \eqref{Eq:ResZeta} that its residue at $z=\zeta_i^\pm$ is equal to $\pm2\cg f_\mp(\zeta_i^\pm)$. Finally, observing that it converges to $0$ when $z$ goes to infinity, we conclude that the partial fraction decomposition of the above function indeed coincides with the last line in \eqref{Eq:LogFlowTwist}. Thus, the flow \eqref{Eq:FlowDefParam} of $z_i$ and $\zeta_i^\pm$ induces the flow \eqref{Eq:FlowTwist2} of $\vp(z)$.

The flow of $\vp(z)$ is related to the ones of the levels $\kay_i$ and $\ell_i$ by Equation \eqref{Eq:FlowTwistLevels}. Using this, one then easily rederives the flows of these levels and thus proves that the flow \eqref{Eq:FlowDefParam} of $z_i$ and $\zeta_i^\pm$ implies
\begin{equation}\label{Eq:FlowKL}
\frac{\dd\kay_i}{\dd\tau} = 0 \;\;\;\;\;\;\; \text{ and } \;\;\;\;\;\;\; \frac{\dd\ell_i}{\dd\tau} = \cg \,f'(z_i) \ell_i = \frac{\cg}{2} \, G^{ii}.
\end{equation}
In particular, this shows the consistency of the RG flow of the level $\kay_i$, as we recover the second equation in \eqref{Eq:RG}.

\paragraph{Flow of the coefficients $\bm{r_i^\pm}$:} In order to check the consistency of the RG flow for the coefficients $\rho_{ij}$, we turn our attention to the coefficients $r_i^\pm$, as the former are expressed in terms of the latter through \eqref{Eq:RhoR}. We will show that their flow coincides with that in \eqref{Eq:FlowRpm}. From the expression \eqref{Eq:LR} of $\ell_i$ in terms of $r^\pm_i$, we have
\begin{equation}
\label{Eq:drSum}
\frac{1}{r_i^+} \frac{\dd r_i^+}{\dd\tau} +  \frac{1}{r_i^-} \frac{\dd r_i^-}{\dd\tau} = \frac{1}{\ell_i} \frac{\dd\ell_i}{\dd\tau}= \frac{\cg}{2} \, G^{ii}\, ,
\end{equation}
where we have used \eqref{Eq:FlowKL} in the second equality. On the other hand, from \eqref{Eq:RPhi} and the definition \eqref{Eq:PhiPm} of the function $\vp_{\pm,i}(z)$, one finds the explicit expression of $r^\pm_i$ in terms of the parameters $z_j$ and $\zeta_j^\pm$ to be 
\begin{equation}
r^\pm_i = \mp \frac{1}{2} \frac{\displaystyle \prod_{j=1}^N \big(z_i - \zeta_j^\pm\big)}{\displaystyle \prod_{j\neq i=1}^N (z_i-z_j)}\, .
\end{equation}
The logarithmic flow of $r^\pm_i$ is thus given by
\begin{equation}
\frac{1}{r^\pm_i} \frac{\dd r^\pm_i}{\dd\tau} = \sum_{j=1}^N \frac{1}{z_i-\zeta^\pm_j}\frac{\dd (z_i-\zeta^\pm_j)}{\dd\tau} -
\sum_{j\neq i=1}^N \frac{1}{z_i-z_j}\frac{\dd (z_i-z_j)}{\dd\tau}\, 
.
\end{equation}
Forming the difference of these logarithmic flows and using \eqref{Eq:FlowDefParam} for the flow of the $z_i$'s and $\zeta_i^\pm$'s, we obtain that
\begin{equation}
\label{Eq:DiffDr.pre}
\frac{1}{r_i^+} \frac{\dd r_i ^+}{\dd\tau} - \frac{1}{r_i^-} \frac{\dd r_i^-}{\dd\tau} =
- 2 \cg \sum_{j=1}^N \left( \frac{f_-(z_i)-f_-(\zeta^+_j)}{z_i-\zeta^+_j} +  \frac{f_+(z_i)-f_+(\zeta^-_j)}{z_i-\zeta^-_j} \right).
\end{equation}
We prove in Appendix~\ref{App:DiffDr} that this can be rewritten as
\begin{equation}\label{Eq:DiffDr}
\frac{1}{r_i^+} \frac{\dd r_i^+}{\dd\tau} - \frac{1}{r_i^-} \frac{\dd r_i^-}{\dd\tau} =
\frac{\cg}{4} \sum_{j=1}^N \left( 4G^{ij} B^j_{\;j} + 2 G^{jj}B^i_{\;j} + \kay_jG^{ij}G^{jj} \right).
\end{equation}
Combining this with \eqref{Eq:drSum}, we finally obtain that
\begin{equation}\label{Eq:FlowR2}
\frac{1}{r_i^\pm}\frac{\dd r_i^\pm}{\dd\tau} = \frac{\cg}{4}G^{ii} \pm \frac{\cg}{8} \sum_{j=1}^N \left( 4G^{ij} B^j_{\;j} + 2 G^{jj}B^i_{\;j} + \kay_jG^{ij}G^{jj} \right) = \Lambda_i^\pm,
\end{equation}
where in the second equality, we used the expression \eqref{Eq:Lambda} for $\Lambda^\pm_i$. This coincides with \eqref{Eq:FlowRpm} found in the previous subsection for the flow of $r^\pm_i$.

\paragraph{Flow of the coefficients $\bm{\rho_{ij}}$:} Let us finally compute the flow of the coefficients $\rho_{ij}$. 
We start with the case $i\neq j$, for which $\rho_{ij}$ is given in \eqref{Eq:RhoR}. The logarithmic flow of $\rho_{ij}$ reads
\begin{equation}
\frac{1}{\rho_{ij}} \frac{\dd\rho_{ij}}{\dd\tau}
 = \frac{1}{r_i^+} \frac{\dd r_i^+}{\dd\tau} + \frac{1}{r_j^-} \frac{\dd r_j^-}{\dd\tau} - \frac{1}{z_i-z_j} \frac{\dd(z_i-z_j)}{\dd\tau}.
\end{equation}
Reinserting in this \eqref{Eq:FlowDefParam} and \eqref{Eq:FlowR2}, we get
\begin{equation}
\frac{1}{\rho_{ij}} \frac{\dd\rho_{ij}}{\dd\tau} = \Lambda_i^+ + \Lambda_j^- - \cg \frac{f(z_i)-f(z_j)}{z_i-z_j}  = \Lambda_i^+ + \Lambda_j^- - \cg \frac{G^{ij}}{2},
\end{equation}
where in the second equality we used \eqref{Eq:GInv} for $G^{ij}$. Comparing the above with \eqref{Eq:ROverRho}, we simply get that
\begin{equation}\label{Eq:FlowRho}
\frac{\dd\rho_{ij}}{\dd\tau} = R^+_{ij}\, , \;\;\;\; \text{ for } i\neq j .
\end{equation}
Thus, the flow of $\rho_{ij}$ ($i\neq j$) induced by the flow \eqref{Eq:FlowDefParam} for $z_i$ 
and $\zeta_i^\pm$ is consistent with the RG flow \eqref{Eq:RG}.

\no Finally, we consider the flow of $\rho_{ii}$ by re-expressing it in terms of the $\rho_{ij}$'s for $j\neq i$, using the definition \eqref{Eq:R} of $r^+_i$:
\begin{equation}
\rho_{ii} = r^+_i - \sum_{j\neq i=1}^N \rho_{ij} + \frac{\kay_i}{2}\, .
\end{equation}
Using 
\eqref{Eq:FlowKL}, \eqref{Eq:FlowR2} and \eqref{Eq:FlowRho}, we then compute the flow of $\rho_{ii}$ as
\begin{equation}
\frac{\dd\rho_{ii}}{\dd\tau} = \Lambda^+_i r^+_i - \sum_{j\neq i=1}^N R^+_{ij}.
\end{equation}
The identity \eqref{Eq:SumR} satisfied by the Ricci tensor then allows us to conclude that
\begin{equation}
\frac{\dd\rho_{ii}}{\dd\tau} = R^+_{ii},
\end{equation}
which achieves the proof of the consistency of the RG flow.

\section{RG flows of non-coupled deformed integrable \texorpdfstring{$\bm\sigma$-models}{sigma}}\label{Sec:deformed}

We proved in Section~\ref{Sec:RGInt} that the RG flow of the integrable $\sigma$-model coupling $N$ copies of the PCM
with WZ terms can be written in a very compact way \eqref{Eq:FlowTwist2} in terms of the twist function $\vp(z)$ of the model, which encodes all of its parameters. It is well known that the model with one copy admits integrable deformations, called the $\lambda$-model~\cite{Sfetsos:2013wia} and the Yang--Baxter (YB) model~\cite{Klimcik:2002zj,Delduc:2014uaa}, which are also characterised by a twist function $\vp(z)$.
Similarly, there exist integrable extensions of the $\lambda$-model with two copies, called the doubly $\lambda$-model~\cite{Georgiou:2017jfi}.
Moreover, the one-loop RG-flow of these deformed models has already been computed in the literature 
\cite{Itsios:2014x,Georgiou:2017jfi,Squellari:2014,Sfetsos:2015nya,Demulder:2017zhz}. 
In this section, we revisit these RG-flows and show that they can be rewritten in terms of the twist function of the models in the exact same way \eqref{Eq:FlowTwist2} as in the undeformed coupled model.

Let us first discuss some common aspects of these deformed models. One of their main characteristics is that their twist function $\vp(z)$ possesses two simple poles $z_\pm$ in the complex plane, as well as two simple zeroes $\zeta_\pm$. In the literature, these zeroes are often fixed to be $+1$ and $-1$, using the freedom of dilation and translation of the spectral parameter $z\mapsto az+b$. For the integrable coupled model studied in this article, this freedom of dilating the spectral parameter has been used in a different way, namely to fix the constant term in the twist function to $-1$ (see the footnote \ref{FootnoteDilation} and Equations \eqref{Eq:Twist} and \eqref{Eq:TwistLevels}). In this section, we will adapt the existing results in the literature to impose the same condition on the twist function of the $\lambda$-model, the doubly $\lambda$-model and the YB model, which will then take a form analogous to \eqref{Eq:TwistLevels}, namely
\begin{equation}
\label{Eq:TwistDef}
\vp(z) = -\frac{(z-\zeta_+)(z-\zeta_-)}{(z-z_+)(z-z_-)} =- \frac{2\kay_+}{z-z_+} - \frac{2\kay_-}{z-z_-} - 1,
\end{equation}
with
\begin{equation}
\kay_\pm=\pm \frac{(z_\pm-\zeta_+)(z_\pm-\zeta_-)}{2(z_+-z_-)}\ .
\end{equation}
As in the case of the undeformed coupled model studied in the rest of this article we define, similarly to \eqref{Eq:SumDiffFpm}, the function
\begin{equation}
\label{Eq:fDef}
\begin{split}
f(z) &= \frac{1}{\vp'(\zeta_+)}\frac{1}{z-\zeta_+} - \frac{1}{\vp'(\zeta_-)}\frac{1}{z-\zeta_-} + \ac
\\
 & =-\frac{1}{\zeta_+-\zeta_-}\left(\frac{(\zeta_+-z_+)(\zeta_+-z_-)}{z-\zeta_+}+\frac{(\zeta_--z_+)(\zeta_--z_-)}{z-\zeta_-}\right)+\ac \, ,
\end{split} 
\end{equation}
with $\ac$ a constant which we do not fix yet. Our goal in this section is then to prove that the RG-flow of
the $\lambda$-model, the doubly $\lambda$-model and the YB model can be recast in terms of the
above functions $\vp(z)$ and $f(z)$ as \eqref{Eq:FlowTwist}, which we recall here for the reader's convenience:
\begin{equation}\label{Eq:FlowTwistDef}
\frac{\dd\;}{\dd\tau} \vp(z) = - \cg \frac{\dd\;}{\dd z} \Big(\vp(z)f(z)\Big)\, .
\end{equation}
This is quite remarkable since the above equation was initially proven for the class of models \eqref{Eq:Action}.
The product $\vp(z)f(z)$ has poles at the points $z_+$ and $z_-$, with corresponding residues $-2\kay_\pm\,f(z_\pm)$, and converges to $-\ac$ when $z$ tends to infinity. This fixes the partial fraction decomposition of $\vp(z)f(z)$ and thus of the right-hand side of Equation \eqref{Eq:FlowTwistDef}.
Using this and inserting \eqref{Eq:TwistDef} into \eqref{Eq:FlowTwistDef}, one finds that
\begin{equation}
\begin{split}
& -\frac{2\kay_+}{(z-z_+)^2} \frac{\dd z_+}{\dd\tau} - \frac{2\kay_-}{(z-z_-)^2} \frac{\dd z_-}{\dd\tau} 
-\frac{2}{z-z_+}\frac{\dd \kay_+}{\dd\tau} - \frac{2}{z-z_-}\frac{\dd \kay_-}{\dd\tau} 
\\
&\qq\qq\qq  = -\frac{2\cg\,\kay_+\,f(z_+)}{(z-z_+)^2} - \frac{2\cg\,\kay_-\,f(z_-)}{(z-z_-)^2}\, .
\end{split}
\end{equation}
Thus it is clear that the RG-flow of the deformed models can be recast as \eqref{Eq:FlowTwistDef} if and only if
\begin{equation}\label{Eq:FlowLZ}
\frac{\dd \kay_\pm}{\dd\tau} = 0 \qquad \text{ and } \qquad \frac{\dd z_\pm}{\dd\tau} = \cg\, f(z_\pm) = \cg\frac{(1+a)\zeta_++(1-a)\zeta_--2z_\mp}{\zeta_+-\zeta_-}\, .
\end{equation}
The first of these conditions states that the $\kay_\pm$'s should be invariants of the RG-flow. Indeed, we will see that they are related to the levels of the WZ terms.
The second condition is the equivalent in the present case of \eqref{Eq:FlowZ} found in the case of the coupled model. 

\no
In the rest of this section we will prove that, indeed, the RG-flow of the $\lambda$-model, the doubly $\lambda$-model and the YB model satisfies the conditions \eqref{Eq:FlowLZ}.

\subsection{The \texorpdfstring{$\bm\lambda$}{lambda}-model}

\paragraph{Definition and twist function:} The action of the $\lambda$-model is~\cite{Sfetsos:2013wia}
\begin{equation}
S_\lambda[g] = S_{\text{WZW},\kay}[g] - \kay \int \dd t\,\dd x \; \text{Tr}\left(\p_+ gg^{-1}\,\frac{1}{\lambda^{-1}-\Ad^{-1}_g} g^{-1}\p_- g \right),
\end{equation}
where the adjoint action $\Ad_g$ is defined by its action $\Ad_g(X)=gXg^{-1}$, reading in 
components $\left(\Ad_g\right)_{ab}=-\Tr\left(T_agT_bg^{-1}\right)$, $\lambda$ and $\kay$ are free parameters (in addition, $4\pi\kay$ is an integer) and $S_{\text{WZW},\kay}[g]$ is the Wess--Zumino--Witten (WZW) action
\begin{equation}
S_{\text{WZW},\kay}[g] = -\frac{\kay}{2} \int \dd t\,\dd x \; \text{Tr}\bigl( g^{-1}\p_+ g g^{-1}\p_- g \bigr) + \kay\, \W g,
\end{equation}
with $\W g$ the WZ term defined in \eqref{action.WZ}.
The $\lambda$-model is integrable~\cite{Sfetsos:2013wia}, in fact in the Hamiltonian sense, as it admits a Lax connection whose spatial component satisfies a Maillet bracket~\cite{Itsios:2014vfa}, from which one reads the twist function $\vp(z)$ of the model~\cite{Hollowood:2015dpa}. Rescaling the spectral parameter so that the constant term in the twist function becomes $-1$, one obtains\footnote{In particular, we read the twist function
from Equations (3.20), (3.26) and (3.27) of~\cite{Hollowood:2015dpa} as
\begin{equation*}
\vp(z)=2\gamma\phi_\epsilon(\gamma z)\,,\quad
k\to\kay\,\pi\,,\quad \gamma=\frac{\lambda}{\kay(1-\lambda^2)}\,.
\end{equation*}}
\begin{equation}\label{Eq:TwistLambda}
\vp(z) = -\frac{2\kay}{z-z_+} + \frac{2\kay}{z-z_-} - 1\, , \;\;\;\;\; \text{ with } \;\;\;\;\; z_\pm = \mp\kay\frac{(\lambda-1)^2}{\lambda}\ .
\end{equation}
Comparing to  \eqref{Eq:TwistDef}, one sees that for the $\lambda$-model, $\kay_\pm=\pm\kay$.
Moreover, one checks that the zeroes of the twist function are
\begin{equation}
\zeta_\pm = \pm \kay \left( \frac{1}{\lambda} - \lambda \right).
\end{equation}
From the definition \eqref{Eq:fDef} of the function $f(z)$, one checks that in the present case it reads
\begin{equation}\label{Eq:fLambda}
f(z) = \frac{\lambda-1}{\lambda+1} \frac{4\kay \,z}{(z-\zeta_+)(z-\zeta_-)} + \ac\, .
\end{equation}

\paragraph{RG-flow:}The RG-flow of the $\lambda$-model was determined in~\cite{Itsios:2014x}. In terms of the parameters $\lambda$ and $\kay$, it reads
\begin{equation}
\frac{\dd \lambda}{\dd\tau} = - \frac{\cg}{\kay} \frac{\lambda^2}{(1+\lambda)^2}\,, \qquad \frac{\dd\kay}{\dd\tau}=0.
\end{equation}
In particular, one sees that the levels $\kay_\pm=\pm\kay$ are invariant under the RG-flow, so that the first condition in \eqref{Eq:FlowLZ} is satisfied. To check the second condition, let us now compute the flow of the poles $z_\pm$. From their expression in \eqref{Eq:TwistLambda}, we simply get
\begin{equation}
\frac{\dd z_\pm}{\dd\tau} = \pm \cg \frac{\lambda-1}{\lambda+1}\, .
\end{equation}
Using the expression \eqref{Eq:fLambda} of the function $f(z)$, one checks that this is equal to $\cg f(z_\pm)$ if and only if we choose the constant term $\ac$ in $f(z)$ to be zero, thus ending the proof that the RG-flow of the $\lambda$-model can be recast as in \eqref{Eq:FlowTwistDef}. Note that, in the case of the integrable coupled $\sigma$-model studied in the previous sections, the constant term $\ac$ in $f(z)$ is unfixed and reflects the freedom of translation of the spectral parameter (see discussion after \eqref{Eq:FlowZ}). In the formulation of the $\lambda$-model considered here, this freedom is in fact fixed by requiring that the zeroes $\zeta_\pm$ are opposite one to another and that forces $\ac=0$.

\subsection{The doubly \texorpdfstring{$\bm\lambda$}{lambda}-model}

\paragraph{Definition and twist function:} Let us now consider the $\lambda$-deformation of the direct product of two current algebras at levels $\kay_1$ and $\kay_2$. This is described by~\cite{Georgiou:2017jfi}
\begin{equation}
\label{double.lambda}
\begin{split}
&S_{\lambda_1,\lambda_2}[g^{(1)},g^{(2)}]=S_{\text{WZW},\kay_1}[g^{(1)}] +S_{\text{WZW},\kay_2}[g^{(2)}]\\
&\qquad -\int\dd t\,\dd x\,\text{Tr}\Big[\begin{pmatrix} R_{1+}  & R_{2+}\end{pmatrix} \binom{\kay_1\Lambda_{21}\lambda_1\Ad^{-1}_{g^{(2)}}\lambda_2 \quad 
\kay_2\lambda_0\Lambda_{21}\lambda_1}{\kay_1\lambda_0^{-1}\Lambda_{12}\lambda_2\quad \kay_2\Lambda_{12}\lambda_2\Ad^{-1}_{g^{(1)}}\lambda_1} \binom{L_{1-}}{L_{2-}}\Big]\,
\end{split}
\end{equation}
and
\begin{equation}
\Lambda_{12}=\left(1-\lambda_2\Ad^{-1}_{g^{(1)}}\lambda_1\Ad^{-1}_{g^{(2)}}\right)^{-1}\,,\quad 
\lambda_0=\sqrt{\frac{\kay_1}{\kay_2}}\, ,
\end{equation}
where $R_{i+}=\partial_+g^{(i)}\,g^{(i)-1}, L_{i-}=g^{(i)-1}\partial_-g^{(i)}$, with $g^{(i)}\in G$, $i=1,2$ and for later use we also introduce
$\kay=\sqrt{\kay_1\kay_2}$.
This model is integrable since its equations of motion can be recast as the zero curvature equations of two independent Lax connections~\cite{Georgiou:2017jfi}. Moreover, the spatial components of these Lax connections satisfy two commuting copies of the Maillet bracket, from which one reads the twist functions $\vp_1(z)$ and $\vp_2(z)$ of the model~\cite{Georgiou:2019plp}. Rescaling the spectral parameters so that the constant term in the twist functions becomes $-1$, we obtain\footnote{In particular, we read the twist functions from Equation (2.29) of~\cite{Georgiou:2019plp} as
\begin{equation*}
(\vp_1(z),\vp_2(z))=
\left(2\gamma_1\vp_\lambda(\gamma_1 z),2\gamma_2\widehat{\vp}_\lambda(\gamma_2 z)\right)\,,
\quad \gamma_{1,2}=\frac{\lambda_{1,2}}{\kay(1-\lambda_{1,2}^2)}\,,\quad
k_{1,2}\to\kay_{1,2}\,.
\end{equation*}}
\begin{equation}
\begin{split}
&\vp_1(z)= -\frac{2\kay\lambda_0^{-1}}{z-z_{1+}} + \frac{2\kay\lambda_0}{z-z_{1-}} - 1\,,\qquad
\vp_2(z)= -\frac{2\kay\lambda_0}{z-z_{2+}} + \frac{2\kay\lambda_0^{-1}}{z-z_{2-}} - 1\,,\\
&z_{1\pm}=\pm\kay\left(2\lambda_0^{\mp1}-\lambda_1-\lambda_1^{-1}\right)\,,\qquad
z_{2\pm}=\pm\kay\left(2\lambda^{\pm1}_0-\lambda_2-\lambda_2^{-1}\right)\,.
\end{split}
\end{equation}
Comparing to \eqref{Eq:TwistDef}, one sees that the levels of the model are
\begin{equation}
\label{double.levels}
\kay_{1+}=\kay\lambda_0^{-1}\,,\quad \kay_{1-}=-\kay\lambda_0\,,\quad
\kay_{2+}=\kay\lambda_0\,,\quad \kay_{2-}=-\kay\lambda_0^{-1}\,.
\end{equation}
Moreover, one checks that the zeroes of the twist functions are
\begin{equation}
\zeta_{1\pm}=\pm\kay\left( \frac{1}{\lambda_1} - \lambda_1 \right)\,,\qquad
\zeta_{2\pm}=\pm\kay\left( \frac{1}{\lambda_2} - \lambda_2 \right)\,.
\end{equation}
From the definition \eqref{Eq:fDef} of the function $f(z)$ we find
\begin{equation}
\label{Eq:fLambdadouble}
\begin{split}
&f_1(z)=-\frac{2\kay}{1-\lambda_1^2}\left(\frac{\lambda_0(1-\lambda_0^{-1}\lambda_1)^2}{z-\zeta_{1+}}
+\frac{\lambda^{-1}_0(1-\lambda_0\lambda_1)^2}{z-\zeta_{1-}}\right)+a_1\,,\\
&f_2(z)=-\frac{2\kay}{1-\lambda_2^2}\left(\frac{\lambda^{-1}_0(1-\lambda_0\lambda_2)^2}{z-\zeta_{2+}}
+\frac{\lambda_0(1-\lambda^{-1}_0\lambda_2)^2}{z-\zeta_{2-}}\right)+a_2\,.
\end{split}
\end{equation}

\paragraph{RG-flow:} The RG-flow of the doubly $\lambda$-model was determined in~\cite{Georgiou:2017jfi}. In terms of the parameters
$\lambda_{1,2}$ and $\kay_{1,2}$, it reads:
\begin{equation}
\begin{split}
&\frac{\dd\lambda_1}{\dd\tau}=-\frac{\cg\lambda_1^2(\lambda_1-\lambda_0)(\lambda_1-\lambda_0^{-1})}{\kay(1-\lambda_1^2)^2}\,,\qquad
\frac{\dd\kay_1}{\dd\tau}=0\, ,
\\
&\frac{\dd\lambda_2}{\dd\tau}=-\frac{\cg\lambda_2^2(\lambda_2-\lambda_0)(\lambda_2-\lambda_0^{-1})}{\kay(1-\lambda_2^2)^2}\,,
\qquad \frac{\dd\kay_2}{\dd\tau}=0\, .
\end{split}
\end{equation}
Let us compare the above RG flows with the two conditions in \eqref{Eq:FlowLZ}.
The first condition is readily satisfied, \textit{i.e.}
\begin{equation}
\frac{\dd \kay_{1\pm}}{\dd\tau} = 0\,,\qquad \frac{\dd \kay_{2\pm}}{\dd\tau} = 0\,,
\end{equation}
as the levels are given by \eqref{double.levels}.
To compare with the second one, we first consider the RG flows on the poles $z_{1\pm}$ and $z_{2\pm}$:
\begin{equation}
\frac{\dd z_{1\pm}}{\dd\tau}=\pm\frac{\cg(\lambda_1-\lambda_0)(\lambda_1-\lambda_0^{-1})}{\lambda_1^2-1}\,,\qquad
\frac{\dd z_{2\pm}}{\dd\tau}=\pm\frac{\cg(\lambda_2-\lambda_0)(\lambda_2-\lambda_0^{-1})}{\lambda_2^2-1}\,.
\end{equation}
Using \eqref{Eq:fLambdadouble}, we find that the second condition in \eqref{Eq:FlowLZ} is also satisfied, \textit{i.e.}
\begin{equation}
\frac{\dd z_{1\pm}}{\dd\tau}=\cg f_1(z_{1\pm})\,,\qquad \frac{\dd z_{2\pm}}{\dd\tau}=\cg f_2(z_{2\pm})\,,
\end{equation}
if we choose the $\ac_i$'s in \eqref{Eq:fLambdadouble} to be
\begin{equation}
a_1=\frac{\lambda_0-\lambda_0^{-1}}{\lambda_1-\lambda_1^{-1}}\,,\qquad
a_2=-\frac{\lambda_0-\lambda_0^{-1}}{\lambda_2-\lambda_2^{-1}}\,,
\end{equation}
ending the proof that the RG-flow of the doubly $\lambda$-model \eqref{double.lambda}
can be recast as \eqref{Eq:FlowTwistDef} or equivalently \eqref{Eq:FlowLZ}.

\subsection{The Yang--Baxter model}

\paragraph{Definition:}  The Yang-Baxter model was originally defined in~\cite{Klimcik:2002zj} and was later generalized to include a WZ term in~\cite{Delduc:2014uaa}. Its action (including a WZ term) takes the form
\begin{equation}\label{Eq:YB}
S_{\text{YB+WZ}}[g] = -\frac{K(1+\eta^2)}{2} \int \dd t\,\dd x \; \text{Tr}\left(  g^{-1}\p_+ g \frac{1-A\, R_g}{1-\eta^2\, R^2_g} g^{-1}\p_- g \right) + \kay\,\W g,
\end{equation}
where $R:\g\mapsto\g$ is the standard Drinfel'd--Jimbo $R$-matrix\footnote{Note that the Yang--Baxter model with WZ term was recently generalized to a larger class of $R$-matrices in~\cite{Hoare:2020mpv}. It would be interesting to check that the present analysis also holds for this more general model.} on $\g$, $R_g = \Ad_g^{-1}  R \,  \Ad_g$ and $K$, $\kay$, $\eta$ and $A$ are constant parameters. The model is integrable if $A$ is related to the other parameters by
\begin{equation}\label{Eq:IntYB1}
A = \eta\sqrt{1-\frac{\kay^2}{K^2(1+\eta^2)}}.
\end{equation}
Using the fact that $R^3=-R$, one rewrites the operator in the action \eqref{Eq:YB} as a polynomial in $R_g$, yielding
\begin{equation}\label{Eq:YB2}
S_{\text{YB+WZ}}[g] =-\frac12 \int \dd t\,\dd x \; \text{Tr}\left(  g^{-1}\p_+ g\, (\alpha + \beta\, R_g + \gamma\, R_g^2) g^{-1}\p_- g \right) + \kay\,\W g\, ,
\end{equation}
with
\begin{equation}
\alpha = K(1+\eta^2), \qquad \beta = K A \qquad \text{ and } \qquad \gamma = K \eta^2\, ,
\end{equation}
following the notations of~\cite{Demulder:2017zhz}. This relation can be inverted as
\begin{equation}
K = \alpha-\gamma, \qquad \eta = \sqrt{\frac{\gamma}{\alpha-\gamma}} \qquad \text{ and } \qquad A = \frac{\beta}{\alpha-\gamma}\, .
\end{equation}
In particular, the integrability condition \eqref{Eq:IntYB1} can be rewritten as
\begin{equation}\label{Eq:IntYB2}
\beta^2 = \frac{\gamma}{\alpha}\bigl(\alpha^2 - \alpha\gamma - \kay^2)\, .
\end{equation}

\paragraph{Twist function:} The YB model admits a Lax representation, as proved in~\cite{Klimcik:2008eq} for the case without WZ term and in~\cite{Delduc:2014uaa} for the case with a WZ term. The corresponding Lax matrix satisfies a Maillet bracket with twist function (see~\cite{Delduc:2013fga} and~\cite{Delduc:2014uaa} for the case with and without WZ terms). In the notations introduced in the previous paragraph, and imposing that its constant term is equal to $-1$, this twist function is given by
\begin{equation}
\vp(z) = \frac{K^2 - z^2}{(z-\kay)^2 + K^2A^2} = \frac{(\alpha-\gamma)^2 - z^2}{(z-\kay)^2 + \beta^2}\, .
\end{equation}
In the notation  \eqref{Eq:TwistDef}, we then read for the poles $z_\pm$, zeroes $\zeta_\pm$ and levels $\kay_\pm$
\begin{equation}
z_\pm = \kay \pm i\beta, \qquad \zeta_\pm = \pm(\alpha-\gamma) \qquad \text{ and } \qquad \kay_\pm = \frac{\kay}{2} \mp i\,\frac{p}{2}\, ,
\end{equation}
where 
\begin{equation}
\label{Eq:h0}
p =\frac{\kay^2-\beta^2-(\alpha-\gamma)^2}{2\beta} = \frac{\kay^2(\alpha+\gamma)-\alpha^2(\alpha-\gamma)}
{\sqrt{4\alpha\gamma(\alpha(\alpha-\gamma)-\kay^2)}}\, ,
\end{equation}
where in the second equality we have eliminated $\beta$ using the integrability condition \eqref{Eq:IntYB2}.
The function $f(z)$ defined in \eqref{Eq:fDef} reads in the present case
\begin{equation}\label{Eq:fYB}
f(z) = \frac{2\kay\alpha(\alpha-\gamma)-z(\kay^2+\alpha^2)}{\alpha\bigl(z^2-(\alpha-\gamma)^2\bigr)} + a\, .
\end{equation}

\paragraph{RG-flow:} The RG-flow of the YB model was determined in~\cite{Squellari:2014,Sfetsos:2015nya}
for the case without WZ term and in~\cite{Demulder:2017zhz} for the case with WZ term.
In terms of the free parameters $\alpha$, $\gamma$ and $\kay$ used in the previous paragraphs, this RG-flow is given by
\begin{equation}
\label{Eq:beta.YB.WZ}
\frac{\dd\alpha}{\dd\tau} = -\frac{\cg}{2} \frac{\kay^2-\alpha^2}{(\alpha-\gamma)^2}\, , \qquad \frac{\dd\gamma}{\dd\tau} =
-\frac{\cg}{2} \frac{\gamma}{\alpha(\alpha-\gamma)^2}(2\alpha\gamma-3\alpha^2+\kay^2) \quad \text{ and } \quad \frac{\dd\kay}{\dd\tau}=0\, .
\end{equation}
Using the integrability condition \eqref{Eq:IntYB2} relating $\beta$ to $\alpha$, $\gamma$ and $\kay$, one gets the RG-flow of $\beta$:
\begin{equation}\label{Eq:FlowBeta}
\frac{\dd\beta}{\dd\tau} = \cg \frac{\beta}{\alpha-\gamma}\, .
\end{equation}
Let us note that the WZ level $\kay$ is an invariant of the RG-flow. Using \eqref{Eq:h0}, one checks that the coefficient $p$ is also RG invariant:
\begin{equation}
\frac{\dd p}{\dd\tau}=0\, .
\end{equation}
Thus, the levels $\kay_\pm=(\kay \mp i p)/2$ are invariant, so that the first condition in Equation~\eqref{Eq:FlowLZ} is satisfied.

\no
The flow of the poles $z_\pm=\kay\pm i\beta$ is obtained simply from the one for $\beta$ in \eqref{Eq:FlowBeta}.
Using the integrability condition \eqref{Eq:IntYB2} and the expression \eqref{Eq:fYB} of $f(z)$,
one checks that this flow coincides with $\cg\,f(z_\pm)$ if one chooses the parameter $a$ to be
\begin{equation}
a = \frac{\kay}{\alpha-\gamma}\, .
\end{equation}
Thus the RG-flow of the YB model with WZ term can be recast in the form \eqref{Eq:FlowTwistDef}.

\section{Conclusion}
\label{Sec:Conclusion}

The main original goal of this work was to study the renormalizability of a class of integrable $\sigma$-models
constructed in~\cite{Delduc:2018hty,Delduc:2019bcl}. This class describes coupled principal chiral models with Wess--Zumino terms, whose action is given by \eqref{Eq:Action}, \eqref{Eq:PhiPm} and \eqref{Eq:RhoK}. To achieve this goal we 
employed geometrical techniques or equivalently heat kernel methods and we proved that they are renormalizable at one-loop order. The derived renormalization group flows take the neat form \eqref{Eq:FlowTwist}, in terms of two functions $\vp(z)$ and $f(z)$: $\varphi(z)$ is the so called twist function, which plays a key r\^ole in the integrability of the models, and $f(z)$ is tightly connected with the twist function. Analyzing \eqref{Eq:FlowTwist} results in the system of renormalization group flow equations \eqref{Eq:FlowDefParam} for the $3N$ defining parameters $z_i$ and $\zeta^\pm_i$ appearing in the expressions \eqref{Eq:PhiPm} and \eqref{Eq:RhoK}.

\no
 A somewhat unexpected important byproduct of our analysis  
is that, as it turns out, \eqref{Eq:FlowTwist} is also applicable  to other classes of integrable $\sigma$-models, whose $r$-matrix  takes the form \eqref{Eq:rtwist}  characterized by a twist function $\vp(z)$. The latter form describes the $\lambda$-deformed model, the doubly $\lambda$-deformed model and the Yang--Baxter model with a Wess--Zumino term. The output of our analysis is that \eqref{Eq:FlowTwist} is in agreement with the one-loop renormalization group flows of these models, which have already been computed in the literature. Hence, we expect that given an integrable $\sigma$-model whose $r$-matrix takes the form \eqref{Eq:rtwist}, and such that all continuous parameters are encoded in the twist function, it is renormalizable at one-loop and its renormalization group flow can be read through \eqref{Eq:FlowTwist}.

\medskip

A natural playground to test this conjecture consists of integrable coupled deformed $\sigma$-models. In the case of $\lambda$-deformations, such integrable coupled models have been introduced in~\cite{Georgiou:2016urf,Georgiou:2017jfi,Georgiou:2018hpd,Georgiou:2018gpe}, their one-loop renormalization group flows have been investigated in~\cite{Georgiou:2017jfi,Georgiou:2018hpd,Georgiou:2018gpe,Sagkrioti:2020mkw}
  and their twist functions computed in~\cite{Georgiou:2019plp}. The most general combination of integrable $\lambda$- and Yang--Baxter deformations of the coupled $\s$-model studied in this article has been constructed in~\cite{Bassi:2019aaf}. Although the twist function of these models is known, their renormalizability has not been studied yet in the literature. It would be interesting to complete the study of the one-loop renormalization group flow of this class of integrable deformed coupled $\sigma$-models and check whether it can be rewritten in terms of their twist function as in Equation \eqref{Eq:FlowTwist}.
Also recently, a nice diagrammatic representation of a general class of integrable $\lambda$-deformations coupled to isotropic principal chiral models and Yang--Baxter $\sigma$-models on group and symmetric cosets was constructed in~\cite{Georgiou:2020wwg}. However, the Hamiltonian integrability in terms of a twist function has not yet been demonstrated for this class of models.

\no 
It would also be interesting to extend our analysis to the Yang--Baxter and $\lambda$-deformed $\sigma$-models on $SU(2)/U(1)$ or more generally on any symmetric coset space, which were constructed in~\cite{Delduc:2013fga} and~\cite{Sfetsos:2013wia,Hollowood:2014rla}, accordingly. The  one-loop renormalization group flows of the aforementioned models have been computed in~\cite{Fateev:1992tk}\footnote{The connection of the Yang--Baxter deformation of $SU(2)/U(1)$~\cite{Delduc:2013fga} with the sausage model~\cite{Fateev:1992tk} 
was noted in~\cite{Hoare:2014pna}. Quite explicitly the fields, parameters and the renormalization group scale 
in equations (3.3), (3.4) and (3.5) of~\cite{Fateev:1992tk}
\begin{equation*}
(X,Y;a,b;t):\quad\text{e}^{Y+iX}\to z\,,\quad a\to 4t(1-\eta^2)\,,\quad b\to 4t(1+\eta^2)\,,\quad t\to-\tau\,,
\end{equation*}
the action matches equation (4.2) of~\cite{Delduc:2013fga} and the renormalization group flows read
\begin{equation*}
\frac{\text{d}\eta}{\text{d}\tau}=2t\eta(1+\eta^2)\,,\quad \frac{\text{d}(t\eta)}{\text{d}\tau}=0\,,
\end{equation*}
see also equations (6.13), (6.14) in~\cite{Sfetsos:2015nya}, when $\zeta=0$.}
and~\cite{Sfetsos:2014jfa,Appadu:2015nfa}, respectively.
These models are integrable in the Hamiltonian sense~\cite{Hollowood:2015dpa,Delduc:2013fga} and although their $r$-matrix does not take the form \eqref{Eq:rtwist}, see~\cite{Delduc:2013fga,Hollowood:2015dpa}, there is a formulation in terms of a twist function (in the language of affine Gaudin models, these correspond to so-called dihedral models~\cite{Vicedo:2017cge}). It would be interesting to investigate what equation replaces \eqref{Eq:FlowTwist} for these models, if any. In addition, it would be also interesting to extend our analysis for the two-parameter integrable deformation of the coset CFT $SU(2)_k\times SU(2)_k/U(1)$~\cite{Georgiou:2016urf}, explicitly constructed in~\cite{Georgiou:2019nbz}, whose one-loop renormalization group flows were computed in~\cite{Georgiou:2017aei}. Finally, it would also be interesting to study the renormalization of the integrable coset models recently constructed in~\cite{Arutyunov:2020sdo}.

\no
In the same lines it would be also worth studying the integrable $\lambda$-deformed non-diagonal coset spaces $G_{k_1}\times G_{k_2}/G_{k_1+k_2}$~\cite{Sfetsos:2017sep}, whose one-loop renormalization group flows were derived in the same work. These models are also integrable in the Hamiltonian sense~\cite{Georgiou:2019plp} and their $r$-matrix takes the form \eqref{Eq:rtwist}. This deformation is expected to be related 
with the integrable bi-Yang--Baxter model~\cite{Klimcik:2014bta}, whose Hamiltonian integrability was demonstrated in~\cite{Delduc:2015xdm}. In particular, these two models share the same $\beta$-functions~\cite{Sfetsos:2017sep,Sfetsos:2015nya} and also twist functions~\cite{Georgiou:2019plp,Delduc:2015xdm} after an analytic continuation (see Equation (3.55) in~\cite{Georgiou:2019plp}). Such a relation is expected since the $\lambda$- and Yang--Baxter deformations are related via Poisson--Lie T-duality
followed by an appropriate analytic continuation ~\cite{Vicedo:2015,Hoare:2015gda,Sfetsos:2015nya,Klimcik:2015gba,Klimcik:2016rov}.

\no
Finally, another possible extension is to study the renormalizability of the bi-Yang--Baxter model with a Wess--Zumino term for a general semi-simple Lie group $G$~\cite{Delduc:2017fib}, which coincides with Lukyanov's four-parameter integrable deformation for the $SU(2)$ case~\cite{Lukyanov:2012zt}. The one-loop renormalization group flow of this model has been worked out in~\cite{Klimcik:2019kkf} for a general semi-simple Lie group $G$ and matches with~\cite{Lukyanov:2012zt} in the $SU(2)$ case. Its Hamiltonian integrability was proven in~\cite{Klimcik:2020fhs} using its interpretation as a $\mathcal{E}$-model, but the derived $r$-matrix cannot be written in terms of a twist function. We expect that this can be achieved using the $G\times G/G_\text{diag}$ coset formulation of the model or the dressing coset one as in~\cite{Delduc:2017fib} and~\cite{Klimcik:2019kkf}, respectively (the notion of dressing 
cosets was introduced in~\cite{Klimcik:1996np} and is also related to the coset construction of~\cite{Sfetsos:1999zm} based on Poisson--Lie  T-dual models).

\no

Beyond studying the various aforementioned examples of integrable $\sigma$-models and checking that their renormalization group flow can be recast in terms of their twist function, it would be interesting to understand whether such a statement can be derived in a model-independent way for a general integrable field theory with twist function, or equivalently a general realization of affine Gaudin model~\cite{Vicedo:2017cge}. It is also worth noticing that the twist function plays an important r\^ole~\cite{Vicedo:2019dej,Delduc:2019c} in the approach to integrable $\sigma$-models through four dimensional Chern--Simons theory recently proposed in~\cite{Costello:2019}. It would therefore be natural to investigate whether these renormalization properties have a natural interpretation in this formalism.

\no
Further possible extensions of our work include studying the higher-loops renormalizability of the integrable coupled $\sigma$-model considered in this article. This question was recently addressed for certain non-coupled integrable deformed $\sigma$-models in~\cite{Hoare:2019ark,Georgiou:2019nbz,Hoare:2019mcc}, showing that their two-loops renormalizability~\cite{Hull:1987yi,Metsaev:1987bc,Metsaev:1987zx,Osborn:1989bu} generically requires adding quantum corrections to the underlying geometry of the target space. It would be interesting to see whether such quantum corrections are also needed in the case of the integrable coupled $\sigma$-model considered here and how these corrections would affect the structure of the geometry of the model put forward in this article (as for instance the various identities obeyed by its metric, B-field and Ricci tensor presented in Subsection~\ref{SubSec:Geom}).

\no
The results of this article bring new evidences of the deep relation between integrable properties of $\sigma$-models and their renormalization group flow. A connection between these two aspects was also recently put forward in~\cite{Hoare:2020fye} for integrable $\sigma$-models on group manifolds and symmetric spaces and their Yang--Baxter and $\lambda$-deformations, relating the renormalization group flow of these models to the existence of a Lax connection for the theory obtained by promoting their parameters to local coupling (depending on the two-dimensional space-time coordinates). It would be interesting to investigate whether a similar relation holds for the integrable coupled $\sigma$-model whose renormalization group flow we determine in the present article.

\subsection*{Acknowledgements}

F. Delduc and S. Lacroix would like to thank Marc Magro and Beno\^it Vicedo. In addition S. Lacroix 
would like to also thank Gleb Arutyunov and Ben Hoare for useful discussions.\\
The research work of S. Lacroix is funded by the Deutsche
Forschungsgemeinschaft (DFG, German Research Foundation) under Germanys Excellence Strategy -- EXC 2121 "Quantum Universe" -- 390833306.\\
The research work of K. Siampos has received funding from the Hellenic Foundation
for Research and Innovation (H.F.R.I.) and the General Secretariat for Research and Technology (G.S.R.T.),
under grant agreement No 15425. 
The research work of K. Sfetsos was supported by the Hellenic Foundation for
Research and Innovation (H.F.R.I.) under the ``First Call for H.F.R.I.
Research Projects to support Faculty members and Researchers and
the procurement of high-cost research equipment grant'' (MIS 1857, Project Number: 16519).\\
K. Sfetsos and K. Siampos would like also to thank the Theoretical Physics Department of CERN for hospitality and
financial support in the initial stages of this work.

\appendix

\section{Geometry and RG-flows}
\label{App:Ge}

\subsection{Generalities}

\paragraph{General $\bm\sigma$-model:} Consider a general $\sigma$-model with local coordinates $y^M$, metric $G_{MN}$ and $B$-field $B_{MN}$. We write its action as
\begin{equation}\label{Eq:LagVielb}
S= \int \dd t\,\dd x\, (G_{AB}+B_{AB})\,\text{e}^A{}_M\text{e}^B{}_N\,\p_+y^M\p_-y^N\, ,
\end{equation}
where we have introduced the vielbeins $\text{e}^A{}_M$ as well as the metric $G_{AB}$ and $B$-field $B_{AB}$ expressed in the tangent space, such that $G_{MN} = \text{e}^A{}_M\text{e}^B{}_N G_{AB}$ and $B_{MN} = \text{e}^A{}_M\text{e}^B{}_N B_{AB}$. The tangent space is spanned by the vectors $\text{e}_A=\text{e}_A{}^M\frac{\p}{\p y^M}$, with $\text{e}_A{}^M$ the inverse vielbeins, with the dual one forms given by $\text{e}^A=\text{e}^A{}_M\,\text{d} y^M$.

\paragraph{One-loop RG flows:}
It is a standard result~\cite{Ecker:1972bm,Honerkamp:1971sh,Friedan:1980jf,Curtright:1984dz,Braaten:1985is,Fridling:1985hc} that the one-loop RG-flow of the metric $G_{MN}$ and $B$-field $B_{MN}$ is given by
\begin{equation}
\label{general.one-loop}
\frac{\dd\;}{\dd\tau} \bigl( G_{MN} + B_{MN} \bigr) = R^+_{MN},
\end{equation}
where $R^+_{MN}$ is the torsionfull Ricci tensor and $\tau=\frac{1}{8\pi}\ln\mu^2$. Equivalently, we may rewrite \eqref{general.one-loop} in the tangent frame $\text{e}^A=\text{e}^A{}_M\text{d}y^M$ as
\begin{equation}
\label{Eq:RGGen}
\frac{\dd\;}{\dd\tau} \bigl( G_{AB} + B_{AB} \bigr) = R^+_{AB}\, ,
\end{equation}
assuming that the $\text{e}^A{}_M$'s are independent of the parameters of the model and hence of the RG scale $\tau$.
The tensor $R^+_{AB}$ can be conveniently written as
\begin{equation}\label{Eq:Ricg}
R^+_{BD}=\partial_A\Omega^+_{BD}{}^A-\partial_B\Omega^+_{AD}{}^A+
\Omega^+_{AE}{}^A\Omega^+_{BD}{}^E-\Omega^+_{ED}{}^A\Omega^-_{AB}{}^E\, ,
\end{equation}
where $\partial_A=\text{e}_A{}^M\frac{\p}{\p y^M}$ and $\Omega^\pm_{AB}{}^C$ are the torsionfull spin-connections
\begin{equation}
\Omega^\pm_{AB}{}^C=\Omega_{AB}{}^C\mp \frac{1}{2} H_{AB}{}^C\, .
\end{equation}
Here $\Omega_{AB}{}^C$ is the standard Levi--Civita connection of the metric and $H_{ABC}$'s are the components of the (totally antisymmetric) field strength of the $B$-field.\footnote{Our conventions for the torsionfull covariant derivative, the torsionfull Riemann and Ricci tensors are
as follows
\begin{equation*} 
\nabla^\pm_AV^B=\partial_AV^B+\Omega^\pm_{AC}{}^BV^C\, ,\quad
[\nabla^\pm_A, \nabla^\pm_B]V^C={{R^\pm_{AB}}^C}_D V^D+(\Omega^\mp_{AB}{}^D-\Omega^\pm_{AB}{}^D)\nabla^\pm_DV^C\, ,\quad
R^\pm_{BD}={{R^\pm_{AB}}^A}_D\, .
\end{equation*}}
The most efficient way to evaluate the $\Omega^\pm_{AB}{}^C$'s is to cast the field equations of the $\sigma$-model in their generic form, namely
\begin{equation}\label{Eq:Conp}
\partial_+(\text{e}^A{}_M\partial_-y^M)+\Omega^+_{BC}{}^A\text{e}^B{}_M\,\text{e}^C{}_N\,\partial_+y^M\partial_-y^N = 0\, ,
\end{equation}
or the parity related one
\begin{equation}\label{Eq:Conm}
\partial_-(\text{e}^A{}_M\partial_+y^M)+\Omega^-_{BC}{}^A\text{e}^B{}_M\,\text{e}^C{}_N\,\partial_-y^M\partial_+y^N= 0\, .
\end{equation}
In what follows we shall apply the above set-up for the $\sigma$-model \eqref{Eq:Action} on $G^N$.

\subsection{Coupled \texorpdfstring{$\bm{\sigma}$}{sigma}-models on \texorpdfstring{$\bm{G^N}$}{GN} and RG flows}

\paragraph{Metric and $\bm{B}$-field:} 
Let us now focus on the $\sigma$-model \eqref{Eq:Action} on $G^N$. The group elements $g^{(i)}$ in each copy of the group $G$ depend on local coordinates $y^{\mu i}$, labelled by pairs of indices $M=(\mu i)$, where $\mu=1,\dots,\dim G$ and $i=1,\dots,N$ with $\dim G$ and $N$ denoting the dimension of the group $G$ and the number of copies, respectively. The tangent indices are pairs $A=(a i)$, where $a=1,\dots,\dim G$ labels an anti-Hermitian basis $\{T_a\}$ of the Lie algebra $\mathfrak{g}$, orthonormalized as $\text{Tr}(T_aT_b)=-\delta_{ab}$.
We can then read the vielbein basis $\text{e}^{a i}{}_{\mu j}$ in terms of the left invariant Maurer--Cartan currents \eqref{leftMC}:
\begin{equation}
\label{Eq:Vielb}
j_\pm^{(i)a}=\text{e}^{ai}{}_{\mu j}\,\p_\pm y^{\mu j}\, ,\qquad
\text{e}^{ai}{}_{\mu j}=-\text{Tr}\Bigg(T_a\,g^{(i)\,-1}\frac{\del g^{(i)}}{\del y^{\mu i}}\Bigg)\, .
\end{equation}
Using the above basis, we can identify the metric and the $B$-field of the model \eqref{Eq:Action} as
\begin{equation}
\label{Eq:FullMetB}
G_{ai\,bj} = \delta_{ab} \,G_{ij} \qquad \text{ and } \qquad B_{ai\,bj} = \delta_{ab}\,B_{ij} + \kay_i\,\delta_{ij}\, W_{ab\,i},
\end{equation}
with $G_{ij}$ and $B_{ij}$ defined as in \eqref{Eq:MetB} and where $W_{ab\,i}=-W_{ba\,i}$ encodes the contribution of the WZ term $\W{g^{(i)}}$, defined in \eqref{action.WZ}, to the $B$-field. By a slight abuse of language, we will also call $G_{ij}$ and $B_{ij}$ the metric and $B$-field of the model.

\paragraph{RG flows for the coupled $\bm\sigma$-model on $\bm{G^N}$:} Our task now is to compute the torsionfull Ricci tensor for the background \eqref{Eq:FullMetB}. Employing the Maurer--Cartan equations \eqref{Eq:Maurer} and the expression \eqref{Eq:Vielb} of the currents $j^{(i)}_\pm$, one rewrites the field equations \eqref{Eq:FieldEq} in the forms \eqref{Eq:Conp} and \eqref{Eq:Conm}, with the torsionfull spin-connections $\Omega^\pm_{AB}{}^{C}$ given by the factorized expression
\begin{equation}\label{Eq:Fact}\Omega^\pm_{ai\,bj}{}^{ck}={f_{ab}}^c\,\omega^{\pm k}{}_{ij}\, .\end{equation}
In the above, ${f_{ab}}^c$ are the structure constants of the Lie algebra $\mathfrak{g}$ and $\omega^{\pm k}{}_{ij}$ are defined as in \eqref{Eq:Omega}. In particular, the $\Omega^\pm_{AB}{}^{C}$'s do not depend on the fields but only on the parameters of the model. Thus, the first two terms in \eqref{Eq:Ricg} vanish. Moreover, the group $G$ is such that the trace in the adjoint representation ${f_{ab}}^b$ vanishes, so that only the last term in \eqref{Eq:Ricg} survives. The torsionfull Ricci tensor then reads
\begin{equation}
R^+_{BD}=-\Omega^+_{ED}{}^A\Omega^-_{AB}{}^E.
\end{equation}
Taking into account (\ref{Eq:Fact}) and the fact that ${f_{db}}^c{f_{ca}}^d = - \cg\, \delta_{ab}$, this leads to
\begin{equation}\label{Eq:RFull}
R^+_{ai\,bj}=-{f_{db}}^c{f_{ca}}^d \omega^{+l}{}_{\;\, kj} \omega^{-k}{}_{\;\, li} =\delta_{ab}R^+_{ij}\, ,
\end{equation}
where $R^+_{ij}$, by a slight abuse of language, will also be called the Ricci tensor and is given by \eqref{RGfinal}, repeated here for the reader's convenience:
\begin{equation}\label{Eq:Ricci}
R^+_{ij} = \cg \sum_{k,l=1}^N \omega^{-k}{}_{\;\, li} \omega^{+l}{}_{\;\, kj}\, ,
\end{equation}
with $\omega^{\pm k}{}_{ij}$ the torsionfull spin connections \eqref{Eq:Omega}.

Let us finally determine the RG flow of the model. The vielbeins $\text{e}^{ai}{}_{\mu j}$ introduced in \eqref{Eq:Vielb} are independent of the parameters of the models, so that \eqref{Eq:RGGen} indeed applies. The metric $G_{ai\,bj}$ and the $B$-field $B_{ai\,bj}$ in the tangent space indices are given by \eqref{Eq:FullMetB}, while the torsionfull Ricci tensor $R^+_{ai\,bj}$ is given by \eqref{Eq:RFull}, containing only terms proportional to $\delta_{ab}$.
Putting altogether into \eqref{Eq:RGGen}, we find that the RG flows 
of $\kay_i$ and $\rho_{ij}$ are given by \eqref{Eq:RG} and \eqref{RGfinal}. Note that the WZ levels $\kay_i$ 
are not renormalized, therefore retaining their topological nature at one-loop order.

\section{Heat-kernel techniques}
\label{appendixheat}

The scope of this appendix is to derive the one-loop RG flows of the $\sigma$-model  \eqref{Eq:Action}  using heat kernel methods. This will provide an independent check of our result. 
We will use a slightly different notation for the coupling  constants of the model which we establish next.
Let us consider $N$ coupled PCMs with WZ terms, whose action reads
\begin{equation}
\label{coupledPCMs}
S=-\frac{1}{8\pi}\sum_{i,j=1}^N\int\dd t\,\dd x\,  E_{ij}\,\text{Tr}(j_+^{(i)} j_-^{(j)})+\sum_{i=1}^Nk_i\,\int {\cal L}^{(i)}_\text{WZ}\, ,
\end{equation}
where the left-invariant Maurer--Cartan currents were defined in \eqref{leftMC} and 
\begin{equation}
\begin{split}
 {\cal L}^{(i)}_\text{WZ} & =\frac{1}{12\pi}\text{Tr}\left(g^{(i)-1}\text{d}g^{(i)}\right)^3
 \\
 &=\frac{1}{4\pi}\text{Tr}\left(g^{(i)-1}\partial_\xi g^{(i)}[g^{(i)-1}\partial_t g^{(i)},g^{(i)-1}\partial_x g^{(i)}]\right)\dd\xi\wedge\dd t\wedge\dd x\,.
 \end{split}
\end{equation}
Matching with \eqref{Eq:Action} yields
\begin{equation}
\label{Translation.of.terms}
\rho_{ij}=\frac{E_{ij}}{8\pi}\,,\quad \kay_i=\frac{k_i}{4\pi}\,.
\end{equation}
We are interested in computing the RG flows of \eqref{coupledPCMs} using the heat kernel technique.
To compute the $\beta$-functions, we need to specify a classical background and compute the quantum fluctuations around it.
The discussion of this section goes along the lines of~\cite{Appadu:2015nfa,Georgiou:2017aei}, see also~\cite{Sagkrioti:2018rwg} for a generalization to generic deformation matrices.

\no
Let us consider the equations of motion for the action \eqref{coupledPCMs}. Varying with respect to the group element
$g^{(i)}$, one finds 
\begin{equation}
\label{eom1}
\sum_{j=1}^N\left\{a_{ij}\partial_-j_+^{(j)}+b_{ij}\partial_+j_-^{(j)}+ E_{ij}[j^{(i)}_+,j^{(j)}_-]- E_{ji}[j^{(j)}_+,j^{(i)}_-]\right\}=0\, ,
\end{equation}
where
\begin{equation}
\label{eom0}
a_{ij}= E_{ji}-K_{ij}\,,\quad b_{ij}= E_{ij}+K_{ij}\,,\quad K_{ij}=k_i\delta_{ij}\, 
\end{equation}
and which is equivalent to \eqref{Eq:FieldEq} with the help of \eqref{Translation.of.terms}.
In addition, we also have the Bianchi identity \eqref{Eq:Maurer}, restated here for the reader's convenience
\begin{equation}
\label{eom2}
\partial_+j_-^{(i)}-\partial_-j_+^{(i)}+[j_+^{(i)},j_-^{(i)}]=0\,.
\end{equation}
We next assume a classical solution $j^{(i)}_\pm$ of \eqref{eom1} and \eqref{eom2} for which the Lagrangian density, when evaluated at that point, is given by \eqref{coupledPCMs}. To carry out the forthcoming calculations, for the case at hand it suffices to use, just as in~\cite{Appadu:2015nfa} and~\cite{Georgiou:2017aei}, a solution such that the currents $j^{(i)}_\pm$ are independent of the space-time point and also commuting.  Varying the equations of motion \eqref{eom1} and the Bianchi identity \eqref{eom2} around this solution, namely $j^{(i)}_\pm+\delta j^{(i)}_\pm$, we obtain the following system of first order differential equations
\begin{equation}
\sum_{j=1}^N\sum_{b=1}^{\text{dimG}}\left(\begin{array}{cc} \delta_{ab}a_{ij}\partial_- +P_{ij|ab}& \delta_{ab}b_{ij}\partial_++Q_{ij|ab}\\
-\delta_{ab}\delta_{ij}\partial_-+R_{ij|ab} & \delta_{ab}\delta_{ij}\partial_++S_{ij|ab}\end{array}\right)
\left(\begin{array}{c}\delta j_+^{(j)b}\\ \delta j_-^{(j)b}\end{array}\right)=0\,,
\end{equation}
where
\begin{equation}
\begin{split}
&P_{ij|ab}=\sum_{m=1}^N E_{im}(\tilde j^{(m)}_-)_{ab}\,\delta_{ij}-(\tilde j^{(i)}_-)_{ab}  E_{ji}\,,\quad
Q_{ij|ab}=\sum_{m=1}^N E_{mi}(\tilde j^{(m)}_+)_{ab}\delta_{ij}-(\tilde j^{(i)}_+)_{ab}  E_{ij}\,,\\
&R_{ij|ab}=(\tilde j^{(j)}_-)_{ab}\delta_{ij}\,,\quad
S_{ij|ab}=-(\tilde j^{(j)}_+)_{ab}\delta_{ij}\,,\quad \left(\tilde j_\pm^{(i)}\right)_{ab}=f_{ab}{}^c\,j_\pm^{(i)c}\,.
\end{split}
\end{equation}
To work out the one-loop effective action, we Wick rotate to Euclidean
and we integrate the fluctuations $\delta j^{(i)}_\pm$'s in the Gaussian path integral, resulting to
\begin{equation}
\label{oneloop}
S_\text{Eucl}^\text{eff}=S+\int\dd t\,\dd x\,\int^\mu\frac{\text{d}^2p}{(2\pi)^2}\ln\det{\cal D}^{-1/2}\,,\quad
 \text{d}^2p=\text{d}p_1\text{d}p_2=p\,\text{d}p\,\text{d}\phi\,,
\end{equation}
where $S$ is the action \eqref{coupledPCMs}, $\mu$ is the energy scale cutoff and the matrix ${\cal D}$ is given by
\begin{equation}
{\cal D}=\left(\begin{array}{cc} \delta_{ab}a_{ij}p_- +P_{ij|ab}& \delta_{ab} b_{ij}p_++Q_{ij|ab}\\
-\delta_{ab}\delta_{ij}p_-+R_{ij|ab} & \delta_{ab}\delta_{ij}p_++S_{ij|ab}\end{array}\right)\,,
\end{equation}
with $p_\pm=p_1\pm i p_2=p\text{e}^{\pm i\phi}.$
To compute the $\beta$-function we need to find the logarithmic dependence in $\mu$ in \eqref{oneloop}.
This can be done by rewriting the determinant as
\be
\label{identity.one-loop.det}
\ln\det{\cal D}=\ln\det C+\text{Tr}\left(C^{-1}F\right)-\frac12\text{Tr}\left(C^{-1}F\right)^2+\dots\,,
\ee
keeping only terms which may diverge when $\mu$ tends to infinity in \eqref{oneloop}, where 
\begin{equation}
C=\left(\begin{array}{cc} \delta_{ab} a_{ij}p_- & \delta_{ab} b_{ij}p_+\\
-\delta_{ab}\delta_{ij}p_- & \delta_{ab}\delta_{ij}p_+\end{array}\right)\,,\qquad
F=\left(\begin{array}{cc} P_{ij|ab} & Q_{ij|ab}\\
R_{ij|ab} & S_{ij|ab}\end{array}\right)\,.
\end{equation}
Next, we evaluate the inverse of $C$ which takes the form
\begin{equation}
C^{-1}=\frac12\left(\begin{array}{cc} \widetilde G^{ij}\frac{\delta_{ab}}{p_-} & -\left(\widetilde G^{-1}b\right)_{ij}\frac{\delta_{ab}}{p_-} \\
\widetilde G^{ij}\frac{\delta_{ab}}{p_+} & \left(\widetilde G^{-1}a\right)_{ij}\frac{\delta_{ab}}{p_+}  \end{array}\right)\,,
\end{equation}
where $\widetilde G_{ij}=\frac12\left( E_{ij}+ E_{ji}\right)$
and $\widetilde G^{ij}=\widetilde G^{-1}_{ij}$.\footnote{Using \eqref{Translation.of.terms}, 
$\widetilde G_{ij}$ is related to $G_{ij}$ in \eqref{Eq:MetB} as 
$\widetilde G_{ij}=8\pi G_{ij}$.} 

The first term in the right-hand side of \eqref{identity.one-loop.det} is field independent, so we can drop it. Moreover, it is easy to check that the second term does not contribute in \eqref{oneloop} upon integration with respect to $\phi$. Turning our attention to the third term, we present here the part with non-vanishing logarithmic contribution in \eqref{oneloop}:
\begin{equation}
\text{Tr}\left(C^{-1}F\right)^2=\frac{c_G}{2p_+p_-}\sum_{i,j,m,n=1}^N{\cal M}_{im}{}^n\left(E;K\right){\cal M}_{jn}{}^m\left(E^T;-K\right)\text{Tr}(j_+^{(i)}j_-^{(j)})+
O\left(\frac{1}{p_+^2},\frac{1}{p_-^2}\right),
\end{equation}
where the omitted non-Lorentz invariant terms vanish after integrating with respect to $\phi$.
In the above, the trace on the left hand side is taken with respect to all indices, whereas in the right-hand side it is taken with respect to the Lie algebra indices only, and we have defined
\begin{equation}
\label{sjfjdlsjhdlsl}
{\cal M}_{im}{}^n(E;K)= E_{in}\left(\widetilde G^{mn}-\widetilde G^{im}\right)+\left(\widetilde G^{-1}(E+K)\right)_{mi}\delta_{ni}\,.
\end{equation}
Finally, we have used $f_{ac}{}^df_{bd}{}^c=-c_G\,\delta_{ab}$.
Putting altogether into \eqref{oneloop},
one finds
\begin{eqnarray}
\label{effective.final}
S_\text{Eucl}^\text{eff}&&=S+\int\dd t\,\dd x\,\int^\mu\frac{\text{d}^2p}{(2\pi)^2}\frac{c_G}{8p_+p_-}\sum_{i,j,m,n=1}^N
{\cal M}_{im}{}^n\left(E;K\right){\cal M}_{jn}{}^m\left(E^T;-K\right)\text{Tr}(j_+^{(i)}j_-^{(j)})\,,\nonumber\\
&&=-\frac{1}{8\pi}\int\dd t\,\dd x\,\sum_{i,j=1}^N  E_{ij}\,\text{Tr}(j_+^{(i)} j_-^{(j)})+
\sum_{i=1}^Nk_i\, \int {\cal L}^{(i)}_\text{WZ}\nonumber\\
&&+\frac{\ln\mu^2}{8\pi}\frac{c_G}{4}\int\dd t\,\dd x\sum_{i,j,m,n=1}^N{\cal M}_{im}{}^n\left(E;K\right){\cal M}_{jn}{}^m\left(E^T;-K\right)
\text{Tr}(j_+^{(i)}j_-^{(j)})\,.
\end{eqnarray}
The one-loop $\beta$-function is derived by demanding that the effective action is independent of the
cutoff scale $\mu$, yielding
\begin{equation}
\label{RGfinal1}
\frac{\text{d} E_{ij}}{\text{d}\ln\mu^2}=\frac{c_G}{4}\sum_{m,n=1}^N{\cal M}_{im}{}^n\left(E;K\right){\cal M}_{jn}{}^m\left(E^T;-K\right)\,,
\end{equation}
where the ${\cal M}_{im}{}^n(E;K)$'s were defined in \eqref{sjfjdlsjhdlsl}.
In the above we have assumed that the elements of the matrix $E$ are much larger than the ones of the matrix $K$, 
such that the one-loop contribution in \eqref{effective.final} is the dominant one. This is illustrated below in the specific example 
of $N$-decoupled PCMs, 
see Eq. \eqref{PCM.beta}.
The levels $k_i$ do not run with the energy scale -- retaining their topological nature at one-loop,
since there is no corresponding counter term in the effective action \eqref{effective.final}.

\paragraph{Properties}

\begin{enumerate}
\item
It can be readily seen that the  $\beta$-function \eqref{RGfinal1} retains its form under the parity transformation \eqref{parity}, \eqref{Translation.of.terms}:
\be
E_{ij}\to E_{ji}\,,\quad k_i\to-k_i\,.
\end{equation}
\item
The derived $\beta$-function \eqref{RGfinal1},  \eqref{sjfjdlsjhdlsl} is equivalent with the analogue one \eqref{Eq:RG}, \eqref{RGfinal}, \eqref{Eq:Omega}.
Indeed using \eqref{Translation.of.terms} one finds
\be
\omega^{-k}{}_{\;\, li}=\frac12{\cal M}_{ik}{}^l\left(E;K\right)\,,\quad \omega^{+l}{}_{\;\, kj}=\frac12{\cal M}_{jl}{}^k\left(E^T;-K\right)
\ee
and with the help of the latter and $\tau=\frac{1}{8\pi}\ln\mu^2$ we can readily show the equivalence of  \eqref{RGfinal1} with \eqref{Eq:RG}.
\item
For $N$-decoupled PCMs $ E_{ij}=\kappa^2_i\delta_{ij}$,  the RG flow equations \eqref{RGfinal1} truncate to
\begin{equation}
\label{PCM.beta}
\frac{\text{d}\kappa_i^2}{\text{d}\ln\mu^2}=\frac{c_G}{4}\left(1-\frac{k_i^2}{\kappa_i^4}\right)\,,
\end{equation}
corresponding to RG flows of $N$-decoupled non-critical WZW models. Note that the range of validity of 
the one-loop result enforces $\kappa_i^2\gg k_i$.
At the IR critical points $k_i=\pm\kappa_i^2$ we retrieve $N$-copies of the WZW model of level $\pm k_i$ respectively.

\end{enumerate}

\section{Two coupled models}
\label{Sec:2coupled}

The scope of this appendix is two study the case with two coupled models, that is
\begin{equation}
\label{dlshfjsksk}
 E_{ij}=
 \left(\begin{array}{cc}
 E_{11}& E_{12}\\  E_{21} &  E_{22}
\end{array}\right)\,,
\end{equation}
using the notations of Appendix \ref{appendixheat}.

\subsection{Integrability}
As we shall show this model is integrable provided that $\mathcal{B}=0$, where
\begin{equation}
\label{constraint.two.copies}
\mathcal{B}=\left( E_{21}+ E_{22}-k_2\right)\left( E_{11}+ E_{21}+k_1\right) E_{12}+\left( E_{11}+ E_{12}-k_1\right)\left( E_{12}+ E_{22}+k_2\right) E_{21}\,.
\end{equation}
Defining the quantities\footnote{Using \eqref{Translation.of.terms}, the $\tilde r^\pm_i$'s are related to the $r_i^\pm$'s in \eqref{Eq:R} as 
$\tilde r^\pm_i=8\pi r^\pm_i$.}
\begin{equation}
\tilde r_i^+=\sum_{j=1}^2E_{ij}-k_i\,,\quad \tilde r_i^-=\sum_{j=1}^2E_{ji}+k_i\,,\quad
\end{equation}
we can rewrite \eqref{constraint.two.copies} as
\begin{equation}
\mathcal{B}=\tilde r_2^+\tilde r_1^-E_{12}+\tilde r_1^+\tilde r_2^-E_{21}\,.
\end{equation}
We then define the parameter
\begin{equation}
\psi=\frac{\tilde r_1^+\tilde r_2^-}{E_{12}}
\end{equation}
and the Lax connection
\begin{equation}
\label{generic.two.copies}
{\mathscr L}_\pm(z)=\pm2\frac{\tilde r_1^\pm(z-\psi)j_\pm^{(1)}+
\tilde r_2^\pm(z+\psi)j_\pm^{(2)}}{z^2\mp2(\tilde r_1^\pm+\tilde r_2^\pm)z\pm2\psi(\tilde r_1^\pm-\tilde r_2^\pm)-\psi^2}\,,\quad z\in\mathbb{C}\,.
\end{equation}
Using \eqref{eom1}, \eqref{eom0} and \eqref{eom2} we can eliminate all the derivatives $\p_\pm j^{(i)}_\mp$ in the Lax curvature
\begin{equation}
\p_+{\mathscr L}_-(z)-\p_-{\mathscr L}_+(z)-[{\mathscr L}_+(z),{\mathscr L}_-(z)]\,,
\end{equation}
expressing it in terms of commutators $[j_+^{(i)},j_-^{(j)}]$ only. It turns out that it is proportional to ${\cal B}$ and as a consequence the connection ${\mathscr L}_\pm(z)$ has zero curvature, provided that ${\cal B}=0$, thus ensuring integrability.
This class of integrable models contains as a subclass the integrable model of~\cite{Delduc:2018hty} for $N=2$ copies, as ${\cal B}$ is vanishing for the parameterization \eqref{Eq:PhiPm}, \eqref{Eq:RhoK}.

\subsection{Renormalizability}

Let us now study the stability of the integrable sector $\mathcal{B}=0$ under the one-loop RG flows  in \eqref{RGfinal1}.
Employing \eqref{constraint.two.copies} and \eqref{RGfinal1} one can show that the constraint $\mathcal{B}=0$ is preserved under the  one-loop RG flows. However, the general system of RG flows is rather complicated even within the integrable sector $\mathcal{B}=0$. Nevertheless, one can still consider a consistent truncation with $ E_{21}=0$, which also enforces that $ E_{22}=k_2$. Then, one may easily check using the action \eqref{coupledPCMs} and the Polyakov--Wiegmann identity~\cite{Polyakov:1983tt} that for the following values 
\begin{equation}
\label{skjskfdhsks}
( E_{11}, E_{12})=(k_1,0),(-k_1,0),(-k_1,2k_1),(-k_1,-2k_2),(k_1+2k_2,-2k_2)\,,
\end{equation}
plus the ones obtained by exchanging 1 and 2 in the above pairs we obtain, respectively, the actions corresponding to 
the exact direct product of WZW conformal field theories (CFTs)
\begin{equation}
G_{k_1}\times G_{k_2}\,,\quad G_{-k_1}\times G_{k_2}\,,\quad G_{-k_1}\times G_{k_1+k_2}\,,\quad G_{k_2}\times G_{-k_1-k_2}\,,\quad
G_{k_1+k_2}\times G_{k_2}\, .
\end{equation}
When both levels are positive these are unitary, otherwise non-unitary.
In the parametric space the unitary and the non-unitary CFTs, are not continuously connected 
as the kinetic term degenerates in between. Unitary CFTs will be shown to be fixed points of the RG flow equations. 

\subsubsection*{An isolated family}

Let us now focus on the RG flow between the unitary CFTs, $G_{k_1}\times G_{k_2}$ and $G_{k_1+k_2}\times G_{k_2}$.
Among the two fixed points, the first one is a stable IR fixed point
whereas the second one is a UV fixed point, which appears to be an unstable one, for generic flows around this point.
To analyze the nature of this fixed point we consider a consistent truncation with $(E_{21},E_{22})=(0,k_2)$. This truncation drastically simplify the system of RG flows
\begin{equation}
\begin{split}
&\frac{\text{d} E_{11}}{\text{d}\ln\mu^2}=-\frac{c_G( E_{11}+k_1)(4k_2^2(k_1- E_{11})+ E_{12}^2(4k_2+ E_{12}))}{( E_{12}^2-4k_2 E_{11})^2}\,,
\\
&\frac{\text{d} E_{12}}{\text{d}\ln\mu^2}=\frac{c_GE_{12}( E_{12}+2k_2)(2k_2(k_1+ E_{11})+ E_{12}(2 E_{11}+ E_{12}))}{( E_{12}^2-4k_2 E_{11})^2}\,.
\end{split}
\end{equation}
Then, we further restrict to the one-parameter family\footnote{\label{Limit.twocopies}
This one-parameter family can be obtained as a singular limit of the integrable models of \eqref{Eq:RhoK},
i.e. the $z_{1,2}$ are colliding. This can be shown by considering a slight modification of
 the one-parameter family \eqref{trunc1}
\begin{equation*}
 E_{11}=k_1+2(k_2-\nu)+\chi\,,\quad  E_{12}=2(\nu-k_2)\,,\quad  E_{21}=0\,,\quad E_{22}=k_2\,,
\end{equation*}
for infinitesimal values of $\chi$. 
The above modification is still integrable as $\mathcal{B}$=0 and takes the form of \eqref{Eq:RhoK}, provided that
\begin{equation*}
\begin{split}
&\zeta^+_1=z_2\,,\quad \zeta^+_2=z_2+\frac{2k_2\chi}{k_2-\nu}\,,\quad \zeta^-_1=z_2+\frac{2\nu^2\chi}{(k_1+k_2)(k_2-\nu)}\,,\\
&\zeta_2^-=z_2-4(k_1+k_2)-2\chi\frac{k_2(k_1+k_2)-2(k_1+k_2)\nu+\nu^2}{(k_1+k_2)(k_2-\nu)}\,,\quad
z_1=z_2+\frac{2\nu\chi}{k_2-x}\,,
\end{split}
\end{equation*}
where we have kept only the linear term around $\chi=0$.}
\begin{equation}
\label{trunc1}
 E_{11}=k_1+2(k_2-\nu)\,,\quad  E_{12}=2(\nu-k_2)\,,\quad E_{21}=0\,,\quad E_{22}=k_2\,,\quad \nu\in[0,k_2]\,,
\end{equation}
leading to
\begin{equation}
\label{betatrunc}
\frac{\text{d}\nu}{\text{d}\ln\mu^2}=-\frac{c_G\,\nu^2(\nu-k_2)(\nu-k_1-k_2)}{2(k_2(k_1+k_2)-\nu^2)^2}\leqslant0\,.
\end{equation}
The one-parameter family of \eqref{trunc1}, describes an integrable interpolation between exact CFTs: at $\nu=0$ we have
the UV CFT $G_{k_1+k_2}\times G_{k_2}$ and at $\nu=k_2$ the IR CFT $G_{k_1}\times G_{k_2}$. The above flow
satisfies the c-theorem~\cite{Zamolo:1986}
\begin{equation}
c_\text{UV}=\frac{2(k_1+k_2)\text{dimG}}{2(k_1+k_2)+c_G}+\frac{2k_2\text{dimG}}{2k_2+c_G}
\quad >\quad
c_\text{IR}=\frac{2k_1\text{dimG}}{2k_1+c_G}+\frac{2k_2\text{dimG}}{2k_2+c_G}\,.
\end{equation}
To understand the above RG flow, we need to rewrite the action \eqref{coupledPCMs} around these points - reading the operator
which drives the theory away from the CFT point.

\paragraph{Around the UV fixed point:}
Around the UV fixed point, the action can be rewritten as\footnote{The renormalization of the class of models in~\eqref{single.lambda.action} was performed
in~\cite{LeClair:2001yp} and~\cite{Georgiou:2017jfi} using CFT and gravitational techniques, respectively. Moreover, for the special case $k_1=0$, where the two WZW models have equal levels, the integrability of~\eqref{single.lambda.action} was studied in~\cite{Bardakci:1996gs} and also follows from the fact that it is a particular limit case of the doubly deformed models of~\cite{Georgiou:2017jfi}. Moreover, from the latter work, it is also integrable in the unequal levels case as well.}
\begin{equation}
\label{single.lambda.action}
S=S_{\text{WZW},k_1+k_2}(g^{(1)})+S_{\text{WZW},k_2}(g) -\frac{\nu}{4\pi} \int\dd t\,\dd x\,\text{Tr}\left(\hat j^{(1)}_+j_-\right)\,,
\end{equation}
where we have introduced $g^{(2)}g^{(1)-1}=g$, $\hat j_+=\partial_+ g g^{-1}$
are the light-cone  right-invariant Maurer--Cartan currents and $j_-=g^{-1}\p_-g$ are the left-invariant ones.
The $\sigma$-model of \eqref{single.lambda.action} as well as its $\beta$-function \eqref{betatrunc}, can be mapped to the ones for the
single $\lambda$-deformed $G_{k_1+k_2}\times G_{k_2}$~\cite{Georgiou:2017jfi} under the following identification of the group elements, parameters and world-sheet coordinates
\begin{equation}
\label{KG.match}
\begin{split}
&(g_1,g_2)\to(g^{(1)},g^{(2)}g^{(1)-1})\,,\qquad  (t,x)\to (2\tau,2\sigma)\,,\\
& (k_1,k_2)\to(k_1+k_2,k_2)\,,\qquad \lambda_0\to\sqrt{\frac{k_1+k_2}{k_2}}\,,\qquad \nu\to\sqrt{(k_1+k_2)k_2}\,\lambda\,.
\end{split}
\end{equation}
The operator driving the perturbation is marginally relevant and
as a consequence the $\beta$ function \eqref{betatrunc} has no linear term once we expand it around $\nu=0$
\begin{equation}
\frac{\text{d} \nu}{\text{d}\ln\mu^2}=-\frac{c_G \nu^2}{2k_2(k_1+k_2)}+O(\nu^3)\,.
\end{equation}

Let us now analyze the field equations of  \eqref{single.lambda.action}. 
These are found using \eqref{eom1}, \eqref{eom0} and \eqref{eom2} for the one-parameter family
\eqref{trunc1} and they read
\begin{eqnarray}
&&\p_+\left((k_1+k_2-\nu)j_-^{(1)}+\nu j_-^{(2)}\right)=0\,,\label{first.single}\\
&&k_2(k_1+k_2-\nu)\p_+j_-^{(1)}+\nu(k_2-\nu)(\p_-j_+^{(1)}-[j_+^{(1)},j_-^{(2)}])=0\,.\label{second.single}
\end{eqnarray}
We can further simplify \eqref{second.single}, using \eqref{first.single}, then \eqref{eom2} to replace the $\p_+j_-^{(2)}$ term and finally the identity
$\left(\Ad^{-1}_{g^{(2)}}\,\p_-\Ad_{g^{(2)}}\right)_{ab}=-f_{ab}{}^c\,j_-^{(2)c}$, yielding
\begin{equation}
\label{second.single.alter}
\p_-\left(\Ad_{g^{(2)}}\left((\nu-k_2)j_+^{(1)}+k_2j_+^{(2)}\right)\right)=0\,.
\end{equation}
Thus, the equations of motion take the form of chiral conserved currents
\begin{equation}
\label{eom.together}
\p_+\left((k_1+k_2-\nu)j_-^{(1)}+\nu j_-^{(2)}\right)=0\,,\quad \p_-\left(\Ad_{g^{(2)}}\left((\nu-k_2)j_+^{(1)}+k_2j_+^{(2)}\right)\right)=0\,,
\end{equation}
and the latter equations are in agreement with Eq.(2.19) of~\cite{Georgiou:2017jfi}, using the map \eqref{KG.match}.

Applying the results of~\cite{Georgiou:2017jfi},
we know that there exists a Lax pair  for \eqref{single.lambda.action}, see Eqs.(2.8) and (3.11) of that work, that is given by
\begin{equation}
\label{Lax.GS}
\begin{split}
&\widehat{{\mathscr L}}_\pm(w)=\frac{2w}{w\mp1} A_\pm\,,\quad 
\p_+\widehat{{\mathscr L}}_--\p_-\widehat{{\mathscr L}}_+=[\widehat{{\mathscr L}}_+,\widehat{{\mathscr L}}_-]\,,\quad w\in\mathbb{C}\,,\\
& A_+=\frac{\nu(k_1+k_2-\nu)}{k_2(k_1+k_2)-\nu^2}\Ad_{g^{(1)}}j_+^{(1)}\,,\quad
 A_-=\frac{\nu(k_2-\nu)}{k_2(k_1+k_2)-\nu^2}\Ad_{g^{(1)}}(j_-^{(1)}-j_-^{(2)})\,.
\end{split}
\end{equation}
Finally, a comment is in order concerning the relation of the Lax connection \eqref{Lax.GS} with the one in \eqref{generic.two.copies}.
Employing \eqref{trunc1} into \eqref{generic.two.copies} we find
\begin{equation}
\label{deg.Lax}
{\mathscr L}_+=0\,,\quad {\mathscr L}_-=4\frac{(k_1+k_2-\nu)j_-^{(1)}+\nu j_-^{(2)}}{4(k_1+k_2)+z}\,.
\end{equation}
As  ${\cal\mathscr L}_+$ vanishes, the zero curvature equation, namely
$ \p_+{\mathscr L}_--\p_-{\mathscr L}_+=[{\mathscr L}_+,{\mathscr L}_-]$,
is equivalent with the first equation in \eqref{eom.together}. Hence, the Lax connection is a degenerate one.
This problem can be solved by considering the model as a limit of a slightly more general but still integrable model, as in footnote
\ref{Limit.twocopies}, whose Lax pair we denote as $\widetilde{\mathscr L}_\pm$.
In that parameterization, taking the limit $\chi\to0$ leads to the degenerate Lax connection \eqref{deg.Lax}.
To avoid this problem, we also rescale the spectral parameter $z\to \chi z$. Putting altogether into \eqref{generic.two.copies}
and taking the limit $\chi\to0$, we find
\begin{equation}
\label{Limit.Lax}
\begin{split}
&\widetilde{\mathscr L}_+(z)=\frac{2(k_2-\nu)}{z(k_2-\nu)+\nu-2k_2}\,j_+^{(1)}\,,\\
&\widetilde{\mathscr L}_-(z)=\frac{(k_1+k_2-\nu)((\nu-k_2)z-\nu)j_-^{(1)}+\nu((\nu-k_2)z+\nu)j_-^{(2)}}{z(k_1+k_2)(k_2-\nu)+\nu(k_1+k_2-2\nu)}\,.
\end{split}
\end{equation}
We will prove that \eqref{Lax.GS} and \eqref{Limit.Lax} are connected with a gauge transformation, namely
\begin{equation}
\widehat{{\mathscr L}}^{\;\,g^{(1)}}_\pm(w)=\Ad^{-1}_{g^{(1)}}\,\widehat{{\mathscr L}}_\pm(w)-j^{(1)}_\pm\,.
\end{equation}
Employing \eqref{Lax.GS} into the latter equation, we find:
\begin{equation}
\begin{split}
&\widehat{{\mathscr L}}^{\;\,g^{(1)}}_+(w)=\frac{2w}{w-1}\frac{\nu(k_1+k_2-\nu)}{k_2(k_1+k_2)-\nu^2}j_+^{(1)}-j_+^{(1)}\,,\\
&\widehat{{\mathscr L}}^{\;\,g^{(1)}}_-(w)=\frac{2w}{w+1}\frac{\nu(k_2-\nu)}{k_2(k_1+k_2)-\nu^2}(j_-^{(1)}-j_-^{(2)})-j_-^{(1)}\,,
\end{split}
\end{equation}
and upon identifying $\widehat{{\mathscr L}}^{\;\,g^{(1)}}_\pm(w)$ in the above expression with $\widetilde{\mathscr L}_\pm(z)$ in \eqref{Limit.Lax} 
we find that the spectral parameters $z$ and $w$ are related by
\begin{equation}
w=\frac{((k_2-\nu)z-\nu)(k_2(k_1+k_2)-\nu^2)}{(k_2-\nu)^2((k_2-\nu)z+3\nu)+k_1(k_2^2z+\nu(2\nu-3k_2)(z-1))}\,.
\end{equation}

\paragraph{Around the IR fixed point:}
Finally, around the IR fixed point the action can be rewritten as
\begin{equation}
S=S_{\text{WZW},k_1}(g^{(1)})+S_{\text{WZW},k_2}(g^{(2)})
+\frac{k_2-\nu}{4\pi}\int\dd t\,\dd x\,\text{Tr}\Big(\hat j_+^{(1)}\Ad_{g^{(1)}}\left(j_{-}^{(1)}-j_{-}^{(2)}\right)\Big)\,.
\end{equation}
In this case, the operator which drives the perturbation has conformal
dimension $2+\Delta_\Ad$, where $\Delta_\Ad$ is the conformal dimension of $\Ad_{g^{(1)}}$, namely
\begin{equation}
\Delta_\Ad=\frac{c_G}{2k_1+c_G}=\frac{c_G}{2k_1}+O\left(\frac{1}{k_1^2}\right)\,.
\end{equation}
The latter is in agreement with the expansion of \eqref{betatrunc} around $\nu=k_2$
\begin{equation}
\frac{\text{d}\nu}{\text{d}\ln\mu^2}=-\frac{c_G}{2k_1}(k_2-\nu)+O((k_2-\nu)^2)\,.
\end{equation}

\section{Inverse metric and \texorpdfstring{$\bm{B}$}{B}-field of the integrable coupled model}
\label{App:GB}

In this appendix, we prove the identities \eqref{Eq:GInv} and \eqref{Eq:B}, for the metric and the $B$-field of the integrable coupled model as defined through Equations \eqref{Eq:MetB} and \eqref{Eq:RhoK}. For that, we start by defining the quantities $G^{ij}$ and $B^i_{\;j}$ by their expressions in Equations \eqref{Eq:GInv} and \eqref{Eq:B} and will prove that $G^{ij}$ is the inverse of the metric and that $B^i_{\;j}$ is equal to the $B$-field with one index raised by $G^{ij}$.

\no
Using the identities \eqref{Eq:FpmId} obeyed by the functions $f_\pm(z)$, one can rewrite $G^{ij}$ as
\begin{equation}\label{Eq:Gt}
G^{ii} = \pm 4 f'_\pm(z_i) \;\;\;\;\; \text{ and } \;\;\;\;\;\; G^{ij} = \pm 4 \frac{f_\pm(z_i)-f_\pm(z_j)}{z_i-z_j}\, , 
\;\; \text{ if } i\neq j\, .
\end{equation}
It was proven in~\cite{Lacroix:2019xeh} that the metric and the $B$-field are related to residues of well-chosen functions. More precisely, one has
\begin{equation*}
\begin{split}
&\rho_{ij} - \frac{\kay_i}{2}\delta_{ij} = \res_{w=z_j} \left(  \res_{z=z_i} \frac{1}{2}\,\frac{\vp_+(z)\vp_-(w)}{z-w} \right)\, ,
\\
&
\rho_{ji} + \frac{\kay_i}{2}\delta_{ij}  = -\res_{w=z_j} \left( \res_{z=z_i} \frac{1}{2}\,\frac{\vp_-(z)\vp_+(w)}{z-w} \right)
\, .
\end{split}
\end{equation*}
On the other hand, the expression of $G^{ij}$ in \eqref{Eq:Gt} coincides with the evaluation of the function 
$$
\pm4\dfrac{f_\pm(z)-f_\pm(w)}{z-w}
$$ at $z=z_i$ and $w=z_j$. To consider contractions of the coefficients $G_{ij}$ and $B_{ij}$ with coefficients $G^{ij}$, it is thus natural to consider the functions
\begin{equation}
\chi^\pm(z,w,t) = 2 \frac{f_\pm(z)-f_\pm(w)}{z-w}\frac{\vp_{\pm}(w)\vp_{\mp}(t)}{w-t}
\end{equation}
and study their evaluation in $z$ and residues in $w$, $t$ at the points $z_i$. Let us then define
\begin{equation}
\chi_{ij}^{\pm} (w) = \res_{t=z_j} \chi^\pm(z_i,w,t)\, ,
\end{equation}
for $i,j\in\lbrace 1,\dots,N\rbrace$. From the definition \eqref{Eq:PhiPm} of the functions $\vp_\pm(z)$ and $\vp_{\pm,i}(z)$, one has
\begin{equation}\label{Eq:PhiOzi}
\vp_\pm(z) = \frac{\vp_{\pm,i}(z_i)}{z-z_i} + \vp'_{\pm,i}(z_i) + O(z-z_i)\, .
\end{equation}
Thus, we get
\begin{equation}\label{Eq:ChiIJ}
\chi_{ij}^{\pm} (w) = 2 \frac{f_\pm(z_i)-f_\pm(w)}{z_i-w} \frac{\vp_{\pm}(w)\vp_{\mp,j}(z_j)}{w-z_j}\, .
\end{equation}
We will now study the residues of $\chi_{ij}^\pm(w)$ at $w=z_k$. It will be useful to distinguish different cases.

\paragraph{Residue for $\bm k$ different from $\bm j$.} Using the expansion \eqref{Eq:PhiOzi} and the fact that the function $\dfrac{f_\pm(z_i)-f_\pm(w)}{z_i-w}$ is regular at $w=z_k$, with value $\pm\frac{1}{4}G^{ik}$, we get
\begin{equation}
\res_{w=z_k} \chi_{ij}^{\pm} (w) = \pm\frac{ G^{ik}}{2} \frac{\vp_{\pm,k}(z_k)\vp_{\mp,j}(z_j)}{z_k-z_j}, \qquad \text{ if } \; k \neq j.
\end{equation}
Then from \eqref{Eq:Rhoij} we obtain that
\begin{equation}\label{Eq:ResChi1}
\res_{w=z_k} \chi_{ij}^+ (w) = G^{ik} \rho_{kj} \qquad \text{ and } \qquad \res_{w=z_k} \chi_{ij}^- (w) =  G^{ik}\rho_{jk}\,  , \qquad \text{ if } \; k \neq j\, .
\end{equation}

\paragraph{Residue for $\bm{k=j}$ but $\bm{i\neq j}$.} First, let us note that
\begin{equation}
\begin{split}
\frac{f_\pm(z_i)-f_\pm(w)}{z_i-w} = \frac{f_\pm(z_i)-f_\pm(z_j)}{z_i-z_j} & + \frac{f_\pm(z_i)-f_\pm(z_j)}{(z_i-z_j)^2}(w-z_j) 
\\
& - \frac{f_\pm'(z_j)}{z_i-z_j}(w-z_j) + O\bigl( (w-z_j)^2 \bigr)\, .
\end{split}
\end{equation}
Using \eqref{Eq:Gt}, this becomes
\begin{equation}
\frac{f_\pm(z_i)-f_\mp(w)}{z_i-w} = \pm \frac{1}{4} G^{ij} \pm \frac{1}{4} \frac{G^{ij}-G^{jj}}{z_i-z_j}(w-z_j) + O\bigl( (w-z_j)^2 \bigr).
\end{equation}
Reinserting this expansion and the one \eqref{Eq:PhiOzi} in Equation~\eqref{Eq:ChiIJ}, we get
\begin{equation*}\begin{split}
\chi_{ij}^\pm(w) = \pm \frac{1}{2}\Big( G^{ij} + \frac{G^{ij}-G^{jj}}{z_i-z_j}(w-z_j) &+ O\bigl( (w-z_j)^2 \bigr) \Big)\times\cr &\times\Big( \frac{\vp_{\pm,j}(z_j)}{w-z_j} + \vp'_{\pm,j}(z_j) + O\bigl( w-z_j \bigr) \Big) \frac{\vp_{\mp,j}(z_j)}{w-z_j}\, .
\end{split}\end{equation*}
The residue of $\chi_{ij}^\pm(w)$ at $w=z_j$ is thus
\begin{equation*}
\res_{w=z_j} \chi_{ij}^{\pm} (w) = \pm \frac{G^{ij}}{2} \vp'_{\pm,j}(z_j)\vp_{\mp,j}(z_j) \pm \frac{1}{2}\frac{G^{ij}-G^{jj}}{z_i-z_j}\vp_{\pm,j}(z_j)\vp_{\mp,j}(z_j)\, , \qquad \text{ if } \; i \neq j\, .
\end{equation*}
From Equations \eqref{Eq:RhoK} and \eqref{Eq:LR}, we have
\begin{equation}\label{Eq:PhiPhi}
\pm \frac{1}{2} \vp'_{\pm,j}(z_j) \vp_{\mp,j}(z_j) = \rho_{jj} \pm \frac{\kay_j}{2} \qquad \text{ and } \qquad \vp_{\pm,j}(z_j) \vp_{\mp,j}(z_j) = - \ell_j\, ,
\end{equation}
hence
\begin{equation}\label{Eq:ResChi2}
\res_{w=z_j} \chi_{ij}^\pm (w) = G^{ij} \left( \rho_{jj} \pm \frac{\kay_j}{2} \right) \mp \frac{\ell_j}{2} \frac{G^{ij}-G^{jj}}{z_i-z_j}\, , \qquad \text{ if } \; i \neq j\, .
\end{equation}

\paragraph{Residue for $\bm{i=j=k}$.} We will need the expansion \eqref{Eq:PhiOzi}, as well as
\begin{equation*}
\frac{f_\pm(z_i)-f_\pm(w)}{z_i-w} = f'_\pm(z_i) + \frac{1}{2} f''_\pm(z_i)(w-z_i) + O\bigl( (w-z_i)^2 \bigr).
\end{equation*}
Starting from Equation~\eqref{Eq:ChiIJ}, we then get
\begin{equation*}\begin{split}
\chi_{ii}^\pm(w) = 2\Big( f'_\pm(z_i) + \frac{1}{2} f''_\pm(z_i)(w-z_i) &+ O\bigl( (w-z_i)^2 \bigr) \Big) \times\cr&\times\Big( \frac{\vp_{\pm,i}(z_i)}{w-z_i} + \vp'_{\pm,i}(z_i) + O\bigl( w-z_i \bigr) \Big) \frac{\vp_{\mp,i}(z_i)}{w-z_i}\, .
\end{split}\end{equation*}
We then extract the residue of this expression at $w=z_i$ as
\begin{equation}
\res_{w=z_i} \chi_{ii}^{\pm} (w) = 2f'_\pm(z_i)  \vp'_{\pm,i}(z_i) \vp_{\mp,i}(z_i) + f''_\pm(z_i) \vp_{\pm,i}(z_i) \vp_{\mp,i}(z_i)\, .
\end{equation}
Using the identities \eqref{Eq:FpmId}, we have
\begin{equation}
\res_{w=z_i} \chi_{ii}^{\pm} (w) = \pm\frac{G^{ii}}{2}  \vp'_{\pm,i}(z_i) \vp_{\mp,i}(z_i) + \left(  \frac{1}{\ell_i} \pm \frac{1}{2} f''(z_i) \right) \vp_{\pm,i}(z_i) \vp_{\mp,i}(z_i)\, .
\end{equation}
From \eqref{Eq:PhiPhi}, we finally get
\begin{equation}\label{Eq:ResChi3}
\res_{w=z_i} \chi_{ii}^\pm (w) = G^{ii}\left( \rho_{ii} \pm \frac{\kay_i}{2} \right) - 1 \mp \frac{\ell_i}{2} f''(z_i)\, .
\end{equation}

\paragraph{Proof of the identities.} With $G^{ij}$ and $B^i_{\;j}$ defined as in \eqref{Eq:GInv} and \eqref{Eq:B}, the residues \eqref{Eq:ResChi1}, \eqref{Eq:ResChi2} and \eqref{Eq:ResChi3} of $\chi_{ij}^\pm(w)$ can be rewritten as
\be
\begin{split}
\res_{w=z_k} \chi_{ij}^+ (w) = G^{ik}\rho_{kj} - \delta_{jk} B^i_{\;j} - \delta_{ij}\delta_{ik}\, , 
\\
\res_{w=z_k} \chi_{ij}^- (w) = G^{ik}\rho_{jk} + \delta_{jk} B^i_{\;j} - \delta_{ij}\delta_{ik}\, .
\end{split}
\ee
We will obtain the desired identities from the property that the sum of all residues of $\chi_{ij}^\pm(w)$ vanishes. For that, we first need to determine whether $\chi_{ij}^\pm(w)$ possesses poles at other points than the $z_k$'s. It is clear from \eqref{Eq:ChiIJ} that, in the complex plane, $\chi_{ij}^\pm(w)$ can have poles at $z_i$, $z_j$ and the poles of $\vp_\pm(w)$ and $f_\pm(w)$. The function $\vp_\pm(w)$ has poles only at the $z_k$'s and thus cannot contribute to poles outside of this set. The function $f_\pm(w)$ has simple poles at the points $\zeta_i^\pm$: however, these are also simple zeroes of the function $\vp_\pm(w)$ in factor. Thus, in the complex plane, $\chi_{ij}^\pm(w)$ can have residues only at the points $z_k$. 
Let us finally study whether it has a residue at infinity. For that, we note that
\begin{equation}
\vp_\pm \left( \frac{1}{u} \right) = 1 + O(u) \qquad \text{ and } \qquad f_\pm(z_i) - f_\pm \left( \frac{1}{u} \right)
 =  O(1)\, .
\end{equation}
Thus, using the expression \eqref{Eq:ChiIJ} of $\chi_{ij}^\pm(w)$, we get
\begin{equation}
\chi_{ij}^\pm \left( \frac{1}{u} \right) = O(u^2)\, ,
\end{equation}
hence proving that $\dfrac{1}{u^2} \chi_{ij}^\pm \left( \dfrac{1}{u} \right)$ is regular at $u=0$, which shows that $\chi_{ij}^\pm$ has no residue at infinity. The vanishings of the sums of residues of $\chi_{ij}^+$ and $\chi_{ij}^-$ then give respectively
\begin{equation}
\sum_{k=1}^N G^{ik}\rho_{kj} = \delta_{ij} + B^i_{\;j} \qquad \text{ and } \qquad \sum_{k=1}^N G^{ik}\rho_{jk} = \delta_{ij} - B^i_{\;j}\, .
\end{equation}
Adding and subtracting the above with the help of \eqref{Eq:MetB}, we find
\begin{equation}
\sum_{k=1}^N G^{ik}G_{kj} = \delta_{ij}\qquad \text{ and } \qquad B^i_{\;j}=\sum_{k=1}^N G^{ik}B_{kj}.
\end{equation}
This ends the proof that $G^{ij}$ in \eqref{Eq:GInv} is indeed the inverse of the metric $G_{ij}$ and that $B^i{}_j$ in \eqref{Eq:B} coincides with the $B$-field with one index raised by $G^{ij}$.

\section{Proof of the identity \eqref{Eq:SumR}} 
\label{App:SumR}

In this appendix, we prove the identity \eqref{Eq:SumR} with the definitions \eqref{Eq:Lambda}, for the torsionfull Ricci tensor of the integrable coupled $\sigma$-model.

\subsection{Summing the Ricci tensor on one index}

Let us consider the torsionfull spin connections $\omega^{- k}{}_{\;\, li}$ and $\omega^{+l}{}_{\;\, kj}$, given in 
Equation~\eqref{Eq:Omega}. Summing over the indices $i$ or $j$ in these connections, we get
\begin{align}
\sum_{i=1}^N \omega^{-k}{}_{\;\, li} &= \frac{G^{kl}}{2} \sum_{i=1}^N \left(\rho_{li} + \frac{\kay_i}{2}\delta_{li} \right)
-\frac{1}{2} \rho^k{}_l + \frac{1}{2} \rho^k{}_l = \frac{1}{2}G^{kl}r^-_l, \\
\sum_{j=1}^N \omega^{+l}{}_{\;\, kj} &= \frac{G^{lk}}{2} \sum_{j=1}^N \left(\rho_{kj} - \frac{\kay_j}{2}\delta_{kj}\right)
- \frac{1}{2} \rho_k{}^l + \frac{1}{2} \rho_k{}^l = \frac{1}{2}G^{kl}r_k^+,
\end{align}
where we used the definition \eqref{Eq:R} of the coefficients $r^\pm_i$. This allows us to compute the sum of the Ricci tensor \eqref{RGfinal} 
on one of its index, namely:
\begin{equation}\label{Eq:SumROmega}
\sum_{i=1}^N R^+_{ij} = \sum_{l=1}^N \Opm - j l r^-_l \;\;\;\;\; \text{ and } \;\;\;\;\; \sum_{j=1}^N R^+_{ij} = \sum_{k=1}^N \Opm + i k r^+_k,
\end{equation}
with
\begin{equation}
\label{Eq:SumROmegas}
\Opm - j l = \frac{\cg}{2} \sum_{k=1}^N G^{kl} \omega^{+l}{}_{\;\, kj} \;\;\;\;\; \text{ and } \;\;\;\;\; \Opm + i k = 
\frac{\cg}{2} \sum_{l=1}^N G^{kl}  \omega^{-k}{}_{\;\, li}.
\end{equation}
Let us focus now on the second sum in \eqref{Eq:SumROmegas}. From the definition \eqref{Eq:Omega} of $ \omega^{-k}{}_{\;\, li}$, one gets
\begin{equation}
\Opm + i k = \frac{\cg}{4} G^{ik}\sum_{l=1}^N G^{kl} (\rho_{li}-\rho_{il}) + \frac{\cg}{4} \sum_{k=1}^N ( G^{kl} )^2 \rho_{il} + \frac{\cg}{8} (G^{ik})^2 \kay_i\, .
\end{equation}
Recalling that $\rho_{li} = 2G_{li}-\rho_{il}$ from \eqref{Eq:MetB}, we then get that
\begin{equation}
\Opm + i k = \frac{\cg}{2} G^{ii} \delta_i^k + \frac{\cg}{4} \sum_{l\neq i=1}^N G^{kl} (G^{kl}-2G^{ik}) \rho_{il} - \frac{\cg}{4} (G^{ik})^2 \left( \rho_{ii} - \frac{\kay_i}{2} \right) .
\end{equation}
Also, from the expression \eqref{Eq:R} of $r^+_i$, we have that
\begin{equation}
\rho_{ii}-\frac{\kay_i}{2} = r^+_i - \sum_{l\neq i=1}^N \rho_{il}\, ,
\end{equation}
hence
\begin{equation}
\Opm + i k = \frac{\cg}{2} G^{ii} \delta_i^k + \frac{\cg}{4} \sum_{l\neq i=1}^N (G^{ik}-G^{lk})^2 \rho_{il} - \frac{\cg}{4} (G^{ik})^2 r^+_i\, .
\end{equation}
So far, the computation holds for any choice of coefficients $\rho_{ij}$ and $\kay_i$ in the action \eqref{Eq:Action}. Let us now consider the integrable model defined by the choice \eqref{Eq:RhoK} of these coefficients. In this case, for $l\neq i$, $\rho_{il}$ can be expressed in terms of $r^+_i$ and $r^-_l$ using Equation~\eqref{Eq:RhoR}, yielding
\begin{equation}\label{Eq:BigOmega}
\Opm + i k = \frac{\cg}{2} G^{ii} \delta_i^k - \frac{\cg}{4} \left( 2 \sum_{l\neq i=1}^N \frac{(G^{ik}-G^{lk})^2 r^-_l}{z_i-z_l} + (G^{ik})^2 \right) r^+_i.
\end{equation}
Thus, for $k \neq i$, we see that $\Opm + i k$ is proportional to $r^+_i$. From Equation~\eqref{Eq:SumROmega}, we then get
\begin{equation}
\sum_{j=1}^N R^+_{ij} = \Lambda^+_i r^+_i\, , 
\end{equation}
with
\begin{equation}
\Lambda^+_i = \Opm + i i - \frac{\cg}{4} \sum_{k\neq i=1}^N \left( 2 \sum_{l\neq i=1}^N \frac{(G^{ik}-G^{lk})^2 r^-_l}{z_i-z_l} + (G^{ik})^2 \right)r_k^+,
\end{equation}
or again, using \eqref{Eq:BigOmega} to express $\Opm + i i$ and relabelling the indices
\begin{equation}\label{Eq:LambdaApp}
\Lambda^+_i = \frac{\cg}{2}G^{ii} - \frac{\cg}{4} \sum_{j=1}^N \left( 2 \sum_{k\neq i=1}^N \frac{(G^{ij}-G^{kj})^2 r^-_k}{z_i-z_k} + (G^{ij})^2 \right)r_j^+.
\end{equation}
A similar reasoning works to find $\Lambda^-_i$, defined by
\begin{equation}
\sum_{j=1}^N R^+_{ji} = \Lambda^-_i \, r^-_i,
\end{equation}
yielding
\begin{equation}
\Lambda^-_i = \frac{\cg}{2}G^{ii} + \frac{\cg}{4} \sum_{j=1}^N \left( 2 \sum_{k\neq i=1}^N \frac{(G^{ij}-G^{kj})^2 r^+_k}{z_i-z_k} - (G^{ij})^2 \right)r_j^-.
\end{equation}

\subsection{Re-expressing \texorpdfstring{$\bm{\Lambda^\pm_i}$}{Lambda pm i}}

Using the identities \eqref{Eq:RhoR} and \eqref{Eq:GInvSym}, one can re-express the coefficient $\Lambda^+_i$ given by \eqref{Eq:LambdaApp}, as
\begin{equation*}\begin{split}
\Lambda^+_i = \frac{\cg}{2}G^{ii} + \frac{\cg}{4} \sum_{j=1}^N \left( \sum_{k\neq i=1}^N (G^{ij}-G^{kj})(G^{ij}-G^{ik}) \rho_{jk}\right.&\left. - (G^{ij})^2 r_j^+ \right) \cr &- \frac{\cg}{2} \sum_{j\neq i=1}^N \frac{(G^{ij}-G^{jj})^2}{z_i-z_j} r^+_j r^-_j\, .
\end{split}\end{equation*}
Combining this with \eqref{Eq:R} and \eqref{Eq:LR}, one obtains
\begin{equation}\begin{split}
\Lambda^+_i = \frac{\cg}{2}G^{ii} + \frac{\cg}{4} \sum_{j=1}^N \left( \sum_{k=1}^N
(G^{ik}G^{jk}-G^{ij}G^{ik}-G^{ij}G^{jk}) \rho_{jk}\right. &\left.+ (G^{ij})^2 \frac{\kay_j}{2} \right) \\&- \frac{\cg}{8} \sum_{j\neq i=1}^N \frac{(G^{ij}-G^{jj})^2}{z_i-z_j} \ell_j\, .
\end{split}\end{equation}
One re-expresses the last term of this equation using the identity \eqref{Eq:GInvToH} as
\begin{equation}
\frac{1}{8} \sum_{j\neq i=1}^N \frac{(G^{ij}-G^{jj})^2}{z_i-z_j} \ell_j = \frac{1}{4} \sum_{j=1}^N (G^{ij}-G^{jj}) \left( B^i_{\;j} + \frac{\kay_j}{2}G^{ij} \right).
\end{equation}
Combining all these together, one obtains a rather simple expression for $\Lambda^+_i$. A similar computation works for $\Lambda^-_i$, as well. In the end, one obtains \eqref{Eq:Lambda}, restated here for the reader's convenience:
\begin{equation}
\Lambda^\pm_i =c_G\left( \frac{G^{ii}}{4} \pm \frac{1}{8} \sum_{j=1}^N \left( 4G^{ij} B^j_{\;j} + 2 G^{jj}B^i_{\;j} + G^{ij}G^{jj}\kay_j \right)\right) .
\end{equation}

\section{Proof of the identity \eqref{Eq:ROverRho}}
\label{App:ROverRho}

In this appendix, we prove the identity \eqref{Eq:ROverRho} obeyed by the torsionfull Ricci tensor. 
Using the definition \eqref{Eq:MetB}, one can rewrite the spin connections \eqref{Eq:Omega} as
\be
\begin{split}
 \omega^{-k}{}_{\;\, li} &= \delta^k_i \delta^k_l + \frac{1}{2}(G^{kl}- G^{ki})\left(\rho_{il} - \frac{\kay_i}{2}\delta_{il} \right) - \frac{\delta_{il}}{2}
\left( \rho_i{}^k - \frac{\kay_i}{2}G^{ik}\right)  , 
\\
\omega^{+l}{}_{\;\, kj} &= \delta^l_j \delta^l_k + \frac{1}{2}(G^{lk}-G^{lj}) \left( \rho_{kj} + \frac{\kay_j}{2}\delta_{kj} \right)
- \frac{\delta_{kj}}{2}\left( \rho^l{}_j+ \frac{\kay_j}{2}G^{lj} \right) .
\end{split}
\ee
Starting from the expression \eqref{RGfinal} of the Ricci tensor $R^+_{ij}$ and re-inserting the above expression for $ \omega^{-k}{}_{\;\, li} $ 
and $\omega^{+l}{}_{\;\, kj}$, one finds that for $i \neq j$
\begin{eqnarray*}
\cg^{-1} R^+_{ij} &=& \frac{G^{ii}+G^{jj}-2G^{ij}}{2} \rho_{ij}
 \\
& & \hspace{10pt} + \frac{1}{4} \sum_{k,l=1}^N \Bigl[ (G^{kl}-G^{ki})(G^{lk}-G^{lj})+(G^{ij}-G^{lj})G^{kl}+(G^{ij}-G^{ik})G^{kl}+G^{ki}G^{lj} \Bigr] \notag \\
& & \hspace{70pt} \times\left( \rho_{il} - \frac{\kay_i}{2}  \delta_{il} \right) \left( \rho_{kj} + \frac{\kay_j}{2}  \delta_{kj} \right) . 
\notag
\end{eqnarray*}
Recalling the definition \eqref{Eq:R} of the quantities $r^\pm_k$, we get
\begin{equation}
\rho_{ii} - \frac{\kay_i}{2} = r^+_i - \sum_{l\neq i=1}^N \rho_{il} \;\;\;\;\; \text{ and } \;\;\;\;\; \rho_{jj} + \frac{\kay_j}{2} = r_j^- - \sum_{k\neq j=1}^N \rho_{kj}
\end{equation}
and thus re-express $R^+_{ij}$ as
\begin{eqnarray}\label{Eq:Rij1}
\cg^{-1} R^+_{ij} &=& \frac{G^{ii}+G^{jj}-2G^{ij}}{2} \rho_{ij} + \frac{(G^{ij})^2}{4} r^+_i r^-_j - \frac{r_i^+}{4} \sum_{k\neq j=1}^N
(G^{ij}-G^{ik})^2 \rho_{kj} \notag \\& &\hspace{30pt}  - \frac{r_j^-}{4} \sum_{l\neq i=1}^N (G^{ij}-G^{lj})^2 \rho_{il}
 + \frac{1}{4} \sum_{\substack{k\neq j=1\\ l \neq i=1}}^N (G^{ij}+G^{kl}-G^{ik}-G^{jl})^2 \rho_{il}\rho_{kj}\, .
\end{eqnarray}
So far, we did not use the expression \eqref{Eq:RhoR} of the coefficients $\rho_{ij}$ which ensures the integrability of the model. From this expression, one gets that for $l \neq i$ and $k\neq j$,
\begin{equation}
\rho_{il} = - \frac{2r^+_i r^-_l}{z_i-z_l} \;\;\;\;\; \text{ and } \;\;\;\;\; \rho_{kj} = - \frac{2r^+_k r^-_j}{z_k-z_j}\, .
\end{equation}
Using this identity, we see that all the terms except the first one in \eqref{Eq:Rij1} are proportional to $r^+_i r^-_j$. Using the fact that the latter is equal to $-\frac{1}{2}(z_i-z_j)\rho_{ij}$, we are then able to factorize $\rho_{ij}$ on the right-hand side of \eqref{Eq:Rij1}. After certain algebraic manipulations, we obtain that
\begin{eqnarray*}
\frac{R^+_{ij}}{\cg\,\rho_{ij}} &=& \frac{G^{ii}+G^{jj}-2G^{ij}}{2}  - \frac{z_i-z_j}{4} \left[ \frac{(G^{ij})^2}{2} + \sum_{k\neq j=1}^N (G^{ik}-G^{ij})X_{i,kj}\, r^+_k 
\notag \right. \\
& &\left. \hspace{30pt} + \sum_{l\neq i=1}^N (G^{ij}-G^{lj})X_{j,il} \,r_l^- \right] + \frac{z_i-z_j}{2} \sum_{\substack{k\neq j=1 \\ l \neq i=1}}^N (X_{k,il} -X_{j,il})(X_{l,kj}-X_{i,kj}) r_k^+ r_l^-\, ,
\end{eqnarray*}
where for any $i,j,k$ with $j\neq k$, we introduced
\begin{equation}
X_{i,jk} = \frac{G^{ij}-G^{ik}}{z_j-z_k}\, ,
\end{equation}
satisfying the property $X_{i,jk}=X_{i,kj}$. In terms of these quantities, the identity \eqref{Eq:GInvSym} can be rewritten as
\begin{equation}
X_{i,jk} = X_{j,ik}\, ,
\end{equation}
for any $k$ distinct from $i$ and $j$. Similarly, the identity \eqref{Eq:GInvToH} gives that
\begin{equation}
\ell_i\,  X_{i,ij} = 2B^j_{\;i} + \kay_i \, G^{ij}\, .
\end{equation}
Moreover, one finds various other identities involving $X_{i,jk}$, such as
\begin{equation}
X_{k,il}X_{l,kj} = \frac{X_{i,kl}(G^{lk}-G^{lj})-X_{j,kl}(G^{lk}-G^{li})}{z_i-z_j}\, , \;\;\;\;\; \text{ if } k \neq j,\, l \neq i,\, k \neq l,\, i\neq j\, .
\end{equation}
We will also need the equations
\begin{equation}
G^{ij} = 2 \sum_{k=1}^N X_{k,ij} \,r_k^+\, ,\qquad G^{ij}= - 2 \sum_{k=1}^N X_{k,ij}\, r_k^-, \;\;\;\;\; \text{ if } \;\; i \neq j\, ,
\label{Eq:GXr}
\end{equation}
the first of which may be proved by considering the holomorphic functions
\begin{equation*}\Psi_{ij}(z)=-\frac{4}{z_i-z_j}\left( \frac{f_+(z)-f_+(z_i)}{z-z_i} - \frac{f_+(z)-f_+(z_j)}{z-z_j }\right) \varphi_+(z)\, ,\quad i\neq j\, .
\end{equation*}
The right-hand side of the first equation in (\ref{Eq:GXr}) is the sum of the residues of the function $\Psi_{ij}$ at the points $z_i, i=1,2,\dots N$, in the complex plane. The left-hand side is the opposite of the residue of the function $\Psi_{ij}$ at infinity. Thus the first equation in (\ref{Eq:GXr})  simply expresses the fact that the sum of residues vanishes. A similar proof holds for the second equation.
Combining all these identities and after a rather long computation, one finds
\begin{equation}
\frac{R^+_{ij}}{\cg\,\rho_{ij}} = \frac{G^{ii}+G^{jj}-2G^{ij}}{4} +
\frac{1}{8} \sum_{l=1}^N \left\{(G^{il}-G^{jl})(4B^l_{\;l}+G^{ll}\kay_l) + 2 G^{ll}(B^i_{\;l}-B^j_{\;l})\right\}\, .
\end{equation}
Using the expression \eqref{Eq:Lambda} of $\Lambda^\pm_i$, we then get
\begin{equation}
\frac{R^+_{ij}}{\rho_{ij}} = \Lambda^+_i + \Lambda^-_j - \frac{\cg}{2} G^{ij}\, ,
\end{equation}
which ends the proof of the identity \eqref{Eq:ROverRho}.

\section{Proof of the identity \eqref{Eq:DiffDr}}
\label{App:DiffDr}

In this appendix, we prove the identity \eqref{Eq:DiffDr} or equivalently, in terms of \eqref{Eq:DiffDr.pre}, the relation
\begin{equation}\label{Eq:IdentityApp}
- 2 \sum_{j=1}^N \left( \frac{f_-(z_i)-f_-(\zeta^+_j)}{z_i-\zeta^+_j} +  \frac{f_+(z_i)-f_+(\zeta^-_j)}{z_i-\zeta^-_j} \right) = \frac{1}{4} \sum_{j=1}^N \left( 4G^{ij} B^j_{\;j} + 2 G^{jj}B^i_{\;j} + \kay_jG^{ij}G^{jj} \right).
\end{equation}

\subsection{Rewriting the left-hand side}

In order to study the left-hand side of \eqref{Eq:IdentityApp} we introduce the function
\begin{equation}\label{Eq:Psi}
\Psi_i(z) = \frac{\bigl(f_+(z)-f_+(z_i)\bigr)\bigl(f_-(z)-f_-(z_i)\bigr)}{z-z_i} \left( \vp'(z) - \frac{\vp(z)}{z-z_i}\right).
\end{equation}
The only factor in this definition which is singular at $z=\zeta_j^\pm$ is $f_\pm(z)-f_\pm(z_i)$. More precisely, we have
that  (see Equation~\eqref{Eq:fpm})
\begin{equation}
\res_{z=\zeta_j^\pm} \bigl( f_\pm(z)-f_\pm(z_i) \bigr)= \frac{1}{\vp'(\zeta_j^\pm)}\, .
\end{equation}
One then computes the residue of $\Psi_i(z)$ at $z=\zeta_j^\pm$ to be
\begin{equation}
\res_{z=\zeta_j^\pm} \Psi_i(z)\,  = \frac{1}{\vp'(\zeta_j^\pm)} \left. \left( \frac{f_\mp(z)-f_\mp(z_i)}{z-z_i} \left( \vp'(z) - \frac{\vp(z)}{z-z_i}\right) \right) \right|_{z=\zeta_j^\pm}.
\end{equation}
Using the fact that $\zeta_j^\pm$ is a zero of $\vp(z)$, we then get
\begin{equation}
\res_{z=\zeta_j^\pm} \Psi_i(z)\,  = \frac{f_\mp(\zeta_j^\pm)-f_\mp(z_i)}{\zeta_j^\pm-z_i}\, .
\end{equation}
Thus, one can rewrite the left-hand side of Equation~\eqref{Eq:IdentityApp} as
\begin{equation}
- 2 \sum_{j=1}^N \left( \frac{f_-(z_i)-f_-(\zeta^+_j)}{z_i-\zeta^+_j} +  \frac{f_+(z_i)-f_+(\zeta^-_j)}{z_i-\zeta^-_j} \right) =
-2 \sum_{j=1}^N \left( \res_{z=\zeta_j^+} \Psi_i(z)\,  + \res_{z=\zeta_j^-} \Psi_i(z)\,  \right)\, .
\end{equation}
A quick study of the behaviour around $z=\infty$ of each term in Equation~\eqref{Eq:Psi} shows that
\begin{equation}
\Psi_i \left(z\right) = O(1/z^4)\, .
\end{equation}
Thus, $\Psi_i(z)$ has no residue at infinity. It is then clear that $\Psi_i(z)$ has poles only at the points $z_j$ and $\zeta_j^\pm$. Using the fact that the sum of its residues vanishes, we then get that
\begin{equation}\label{Eq:SumResZ}
- 2 \sum_{j=1}^N \left( \frac{f_-(z_i)-f_-(\zeta^+_j)}{z_i-\zeta^+_j} +  \frac{f_+(z_i)-f_+(\zeta^-_j)}{z_i-\zeta^-_j} \right) =
2 \sum_{j=1}^N \res_{z=z_j} \Psi_i(z)\, .
\end{equation}
In the next subsections, we will compute the residues of $\Psi_i(z)$ at the points $z_j$, to 
appropriately re-express the right-hand side of \eqref{Eq:SumResZ}.

\subsection{Residue at \texorpdfstring{$\bm{z_i}$}{zi}}

In this subsection, we compute the residue of $\Psi_i(z)$ in \eqref{Eq:Psi}  at $z=z_i$. 
From the partial fraction decomposition \eqref{Eq:TwistLevels} of $\vp(z)$, we get that
\begin{equation}
(z-z_i)\vp'(z) - \vp(z) = -\frac{3\ell_i}{(z-z_i)^2} + \frac{4\kay_i}{z-z_i} + O\bigl((z-z_i)^0\bigr)\, .
\end{equation}
Moreover, we have
\begin{equation*}
\frac{f_\pm(z)-f_\pm(z_i)}{z-z_i} = f_\pm'(z_i) + \frac{1}{2} f_\pm''(z_i) (z-z_i) + O\bigl((z-z_i)^2\bigr)\, .
\end{equation*}
Inserting the above into the definition \eqref{Eq:Psi} of $\Psi_i(z)$ we find, after some algebra,
\begin{align*}
\Psi_i(z) &=  - \frac{3\ell_i f_+'(z_i) f_-'(z_i)}{(z-z_i)^2} + \frac{8 \kay_i f_+'(z_i) f_-'(z_i) - 3\ell_i f_+'(z_i) f_-''(z_i) - 3\ell_i f_+''(z_i) f_-'(z_i)}{2(z-z_i)} \notag \\
&\hspace{350pt}+ O\bigl((z-z_i)^0\bigr)\, ,
\end{align*}
hence
\begin{equation}
\res_{z=z_i} \Psi_i(z) \,   = 4 \kay_i f_+'(z_i) f_-'(z_i) - \frac{3\ell_i}{2} f_+'(z_i) f_-''(z_i) - \frac{3\ell_i}{2} f_+''(z_i) f_-'(z_i)\, .
\end{equation}
Using the identities \eqref{Eq:FpmId}, we then get that
\begin{equation}
\res_{z=z_i} \Psi_i(z) \,
= \frac{3\ell_i}{4} f'(z_i) f''(z_i) - \kay_i\, f'(z_i)^2 = \frac{3}{2} f'(z_i) \left( \frac{\ell_i}{2}f''(z_i) - \kay_i\,f'(z_i) \right) + \frac{\kay_i}{2} f'(z_i)^2 .
\end{equation}
Using the expressions \eqref{Eq:GInv} and \eqref{Eq:B} of $G^{ii}$ and $B^i_{\;i}$, we finally obtain
\begin{equation}
\res_{z=z_i} \Psi_i(z) \,   = \frac{3}{4} G^{ii} B^i_{\,i} + \frac{\kay_i}{8} \left( G^{ii}\right)^2 .
\end{equation}
Let us note that this can be rewritten as
\begin{equation}\label{Eq:ResPsi1}
\res_{z=z_i} \Psi_i(z) \,   = \frac{1}{2} G^{ij} B^j_{\,j} + \frac{1}{4} G^{jj} B^i_{\,j} + \frac{\kay_j}{8} G^{ij} G^{jj}\,, \;\;\;\; \text{ for } \;\;\;\; j=i .
\end{equation}

\subsection{Residues at \texorpdfstring{$\bm{z_j}$}{zj} for \texorpdfstring{$\bm{j\neq i}$}{j not i}}
\label{App:Res3}

In this subsection, we compute the residue of $\Psi_i(z)$ at $z=z_j$ for $j$ different than $i$. Let us study the behavior of the different terms forming $\Psi_i(z)$ around $z=z_j$. For instance, we have
\begin{equation*}
\frac{1}{z-z_i} = -\frac{1}{z_i-z_j} - \frac{1}{(z_i-z_j)^2} (z-z_j) - \frac{1}{(z_i-z_j)^3}(z-z_j)^2 + O\bigl((z-z_j)^3\bigr)\, .
\end{equation*}
Moreover, using the identities \eqref{Eq:FpmId}, one obtains
\begin{equation*}\begin{split}
f_\pm(z)-f_\pm(z_i) = \pm \frac{1}{2} \Big( f(z_j)-f(z_i) + f'(z_j) (z-z_j) + \frac{1}{2}\left( f''(z_j) \pm \frac{2}{\ell_i} \right)& (z-z_j)^2 \cr&+ O\bigl((z-z_j)^3\bigr) \Big).
\end{split}\end{equation*}
From the partial fraction decomposition \eqref{Eq:TwistLevels} of $\vp(z)$, one gets
\begin{equation*}\begin{split}
\vp'(z) - \frac{\vp(z)}{z-z_j} = - \frac{2\ell_j}{(z-z_j)^3} + \left( \frac{\ell_j}{z_i-z_j} + 2\kay_j \right) \frac{1}{(z-z_j)^2} + \left( \frac{\ell_j}{(z_i-z_j)^2} -\right. &\left.\frac{2\kay_j}{z_i-z_j} \right) \frac{1}{z-z_j}
\cr & + O\bigl((z-z_j)^0\bigr).
\end{split}\end{equation*}
Combining all these results, one computes the residue of $\Psi_i(z)$ at $z=z_j$, yielding that
\begin{align*}
\res_{z=z_j} \Psi_i(z) &= \frac{1}{4} \left( \frac{\ell_j}{(z_i-z_j)^2} - \frac{2\kay_j}{z_i-z_j} \right) \frac{(f(z_i)-f(z_j))^2}{z_i-z_j}  \\
&  + \frac{1}{4}\left( \frac{\ell_j}{z_i-z_j} + 2\kay_j \right) \left( \frac{(f(z_i)-f(z_j))^2}{(z_i-z_j)^2} - \frac{2(f(z_i)-f(z_j))f'(z_j)}{z_i-z_j} \right) \\
&- \frac{\ell_j}{2} \left( \frac{(f(z_i)-f(z_j))^2}{(z_i-z_j)^3} - \frac{2(f(z_i)-f(z_j))f'(z_j)}{(z_i-z_j)^2} + \frac{f'(z_j)^2 - \bigl( f(z_i) - f(z_j) \bigr)f''(z_j)}{z_i-z_j} \right).
\end{align*}
From the expressions \eqref{Eq:GInv} and \eqref{Eq:B} for $G^{ij}$ and $B^i_{\;j}$ and after a few manipulations, one re-expresses the above residue as
\begin{equation}\label{Eq:ResPsi2}
\res_{z=z_j} \Psi_i(z) \,   = \frac{1}{2} G^{ij} B^j_{\,j} + \frac{1}{4} G^{jj} B^i_{\,j} + \frac{\kay_j}{8} G^{ij} G^{jj}\,, \;\;\;\; \text{ for } \;\;\;\; j\neq i\, .
\end{equation}
Re-inserting the expressions \eqref{Eq:ResPsi1} and \eqref{Eq:ResPsi2} of the residues of $\Psi_i(z)$ at $z=z_i$ and $z=z_j$ ($j\neq i$) in \eqref{Eq:SumResZ}, we obtain the desired identity \eqref{Eq:IdentityApp}, as announced.

\bibliographystyle{unsrt}

\end{document}